\documentclass{statsoc}

\usepackage{amsmath,amssymb}
\graphicspath{{figures/}} 

\usepackage{rotating}
\usepackage{multirow}
\usepackage{geometry,setspace}
\usepackage{color}
\usepackage{hyphenat} 
\usepackage{enumitem}
\usepackage{bm}

\usepackage{etoolbox}
\AtBeginEnvironment{table}{\tiny}

\makeatletter
\patchcmd{\@makecaption}
{\parbox}
{\advance\@tempdima-\fontdimen2} 
{}{}
\makeatother

\usepackage[caption = false]{subfig} 
\usepackage{enumerate}
\usepackage{comment}
\usepackage{natbib}
\usepackage[section]{placeins}
\usepackage{float}

\newcommand{\mY}{\mbox{\textbf{Y}}}
\newcommand{\bY}{\mbox{\bm{$Y$}}}
\newcommand{\mX}{\mbox{\textbf{X}}}

\newcommand{\argmin}{\arg\!\min}
\newcommand{\bpsi}{\mbox{\boldmath{$\psi$}}}

\DeclareMathOperator*{\E}{\mathbb{E}}


\newcommand{\beginsupplement}{%
	\setcounter{table}{0}
	\renewcommand{\thetable}{S\arabic{table}}%
	\setcounter{figure}{0}
	\renewcommand{\thefigure}{S\arabic{figure}}%
}

\usepackage{hyperref}

\title[Block bootstrap for microbiome data]{The Block Bootstrap Method for Longitudinal  Microbiome Data}

\author[Pratheepa Jeganathan {\it et al.}]{Pratheepa Jeganathan}
\address{Statistics Department,
	 CA 94305 Stanford,	USA.}
\email{pjeganat@stanford.edu}
\author{Benjamin J. Callahan}
\address{Department of Population Health and Pathobiology, NC State University, Raleigh NC, USA}
\author{Diana M. Proctor}
\address{Department of Infectious Disease, Stanford University School of Medicine, Stanford, California, USA}
\author{David A. Relman}
\address{Department of Microbiology \& Immunology, Stanford University School of Medicine, Stanford, California, USA, Department of Medicine, Stanford University School of Medicine, Stanford, California, USA, Infectious Diseases Section, Veterans Affairs Palo Alto Health Care System, Palo Alto, California, USA}
\author[Pratheepa Jeganathan {\it et al.}]{Susan P. Holmes}
\address{Statistics Department,
	 CA 94305 Stanford,	USA.}
 \email{susan@stat.stanford.edu}    

\begin{document}
	\maketitle
	\begin{abstract}
		Microbial ecology serves as a foundation for a wide range of scientific and biomedical studies. Rapidly evolving high-throughput sequencing technology enables the comprehensive search for microbial biomarkers using longitudinal experiments. Such experiments consist of repeated biological observations from each subject over time and are essential in accounting for the high between-subject and within-subject variability. 

Unfortunately, many of the statistical tests based on parametric models rely on correctly specifying temporal dependence structure which is un- available in most microbiome data.

In this paper, we propose an extension of the nonparametric bootstrap method that enables inference on these specific types of longitudinal data. The proposed MBB method accounts for within-subject dependency by using overlapping blocks of repeated observations within each subject to draw valid inferences based on approximately pivotal statistics. Simulation studies show that compared to tests that presume independent samples (PIS) or merge-by-subject (MBS), our method has high true positive rates and small values of false positive rates for practical FDR cut-off values. 

In this paper, we illustrated the MBB method using three different pregnancy datasets and an oral microbiome data. We provide an open-source R package \href{https://github.com/PratheepaJ/bootLong}{\textit{bootLong}} to make our method accessible and the study in this paper reproducible. 

	\end{abstract}

	\section{Introduction} 
	High-throughput marker-gene and metagenomics sequencing (MGS) of environmental DNA provide a cost-effective and culture-independent description of microbial communities \citep{LSGJK2012, YRMTDMCMHBA2012}. The adoption of MGS methods has transformed the study of the complex microbial communities, or ``microbiomes", that inhabit environments ranging from the human body to waste-water treatment plant on the hyper-saline desert \citep{T2007,Gilbert2010}.

Microbial communities vary significantly over time and space, and among subjects, in studies of host-associated microbiomes \citep{E2005, T2007, LSGJK2012,  FCHJJDCLJ2014, P2017}. Microbiome researchers have increasingly adopted longitudinal experimental designs consisting of repeated biological observations from each subject \citep{SAS2013, FCHJJDCLJ2014, DCM2015, F2017, Proctor2018} for purposes such as the detection of microbial biomarkers of disease. Longitudinal designs help experimenters overcome some of the difficulties caused by the high temporal and subject-to-subject variability of microbiomes, and allow subjects to be used as their own `controls' \citep{CW1983, D2002}.

Currently, the analysis of longitudinal microbiome data is limited by the lack of appropriate statistical methods that account for both the temporal dependency due to repeated observations, and specific characteristics such as sparsity and high dimensionality of microbiome data \citep{X2017}. These data typically have fewer samples than the number of measured bacteria, and technical variability due to batch effects and unequal sequencing depth of samples can be comparable to the levels of biological variability. Moreover, mean-variance dependencies and variation in within-subject correlation profiles introduce heteroscedasticity \citep{MH2014}.

One standard statistical analysis in microbiome studies is differential abundance testing where the association of bacterial abundances with two or more experimental states or conditions is tested in a set of independent samples (e.g. health and disease). Recent methods have been developed to account for the sparsity and heteroscedasticity. In particular, the similarity between testing for differential gene expression and differentially abundant bacteria was leveraged by \citet{MH2014} in their proposal to use gamma-Poisson mixtures and adapting the R package DESeq2 \citep{L2014}  to the microbiome context. This was shown to be effective in accounting for the technical variability of microbiome measurements due to unequal sequencing depth. In addition, the researchers using these methods typically presume independent samples.  

However, presuming independent samples (PIS) leads to high false positive rates and findings that cannot be replicated when applied to longitudinal data. Parametric mixture models that account for  dependencies can only be applied if one can specify the temporal dependence structure within a subject \citep{S2016}. 

For longitudinal data with a response variable, either continuous or discrete, \citet{D2002} proposed generalized marginal, mixed effect, or transitional model-based asymptotic inferences. These generalized linear models can be extended to longitudinal microbiome data. A recent negative binomial mixed effects model (NBMEM) approach for detecting differentially expressed genes used permutation testing \citep{S2016}. Although the modeling accounts for dependency due to repeated observations as a random effect, permutation testing for inference is only valid for exchangeable or conditionally exchangeable observations. In longitudinal microbiome data, repeated observations within a subject have different autocorrelations that prevent these types of exchangeability. Another NBMEM approach accounts for the temporal dependency by a normal distribution\citep{zhang2017}. The same normal distribution assumption for all the subjects hugely impacts the false positive rate and may not be valid for microbiome data in practice due to between-subject variability.

A simple conservative alternative (used for instance in \citet{DCM2015}) is to merge abundance measurements by subject, thereby collapsing the temporal dependence structure. Merging-by-subject (MBS) refers to averaging abundances within-subject. In the MBS method, temporal dynamics and the individual values of multiple within-subject observations are ignored, thereby eliminating within-subject dependency at the cost of sample size and power.

To overcome the issues stated above, we propose a moving block bootstrap (MBB) method with an optimal block size step. The bootstrap  is a computationally intensive method which enables  inferences when analytic derivations are unavailable for the sampling distributions of estimators. The original bootstrap method \citep{E1979} developed for independent and identical variables has been widely extended to many applications where the samples are dependent. Extension approaches have included either by modeling the dependency structure and resampling residuals (parametric bootstrap) or by changing the resampling mechanism to account for these dependencies (nonparametric bootstrap) \citep{ET1994,DH1997, DH1993,HHJ1995,berkowitz2000,L1999,Politis2004}. 

The MBB accounts for autocorrelation within a time series by a blocking and resampling procedure \citep{LB1992, DH1993, L1999} in which blocks of temporally contiguous observations are constructed and resampled with replacement to make bootstrap realizations that approximate the distribution of the chosen statistic.  Carlstein's (\citeyear{carlstein1986})  non-overlapping methods and K\.{u}nsch's (\citeyear{kunsch1989}) and Liu \& Singh's (\citeyear{LS1992}) overlapping are two such blocking procedures for time series \citep{HHJ1995}. Because \citet{L1999} showed that the latter produces the least mean squared error for most statistics and the number of repeated observations are small in biomedical experiments, the overlapping block bootstrap is preferred here.
 
Crucial to the MBB method is identifying the optimal block size that best captures the temporal autocorrelation of the data. Intuitively, this means that we must evaluate  the ``effective sample number of observations" of a time series \citep{Box1976}. If we have $q$ dependent observations, the effective size will be smaller than $q$ but how small depends on the level of correlation between observations. At one extreme if the correlation is almost perfect, then we cannot do better than the MBS procedure. As the dependency decreases we will come closer to the PIS model. 

In time series analysis, \citet{HHJ1995} proposed an empirical subsampling method to choose an optimal block size. First, this method minimizes the average squared difference between the bootstrap estimator (bias, variance, one-sided and two-sided probabilities) computed with the full time series and the one constructed from many subsets of overlapping subseries (hereafter, subsamples). Finally, the optimal block size for the full time series is obtained by using the optimal block size for the subseries and a formula in \cite{HHJ1995}. Here we propose a modified empirical subsampling method that identifies the optimal block size by minimizing the mean squared error (MSE) of a block bootstrap estimator of two-sided probability.

In addition to the proposed MBB and subsampling, we describe appropriate pre-processing and transformation procedures, as well as the construction and interpretation of statistical tests.  Microbiome data consist of the frequency of higher resolution amplicon sequence variants (ASVs) \citep{CMRHJH2016} in each biological sample. As a first step, we recommend eliminating ASVs with very low counts; this minimizes the number of multiple hypotheses tested.  The log transformations are often used in microbiome studies. However, they are undefined in the presence of zero abundances and lead to a lack of power in detecting changes between the lower abundance ASVs. As the data often follow a gamma-Poisson mixture model \citep{MH2014}, we can use the optimal variance-stabilizing transform for that distribution (arcsinh as originally proposed by \citet{Anscombe1948}). Then, we choose a pivotal statistic to measure the discrepancy between the null hypothesis and sample information. Finally, both confidence intervals or p-values can be used to determine statistical significance at the desired false discovery rate (FDR) cut-off value.

In section \ref{method}, we describe  the unifying  MBB method for differential abundance analysis. We then provide the subsampling algorithm and the method used for choosing the optimal block size. In section \ref{simulation}, we demonstrate the MBB method on the specific problem of inferring differentially abundant bacteria between the case and control cohorts using synthetic and real longitudinal microbiome data and compare the MBB method with the MBS and PIS approaches. In section \ref{con}, we conclude with perspectives for future work.

	\section{Method} \label{method}
	\subsection{Moving Block Bootstrap Method for Microbiome Data}

In this section, we illustrate how a unifying nonparametric methodology enables inference in the presence of an unknown order of within-subject weak dependence in longitudinal ASVs abundances. 
		 
\begin{table}
	\caption{\label{tab01}Count matrix $ \mY \in \mathbb{R}^{m \times N}$}
	\fbox{%
		\begin{tabular}{l|l l l | l | l l l |l | l l l}
			ASV & \multicolumn{3}{|c|}{Subject 1}& $\cdot \cdot$& \multicolumn{3}{c|}{Subject $j$} & $\cdot \cdot$& \multicolumn{3}{c|}{Subject $n$}\\
			\hline
			$ASV_{1}$ & $Y_{111}$ &  $\cdot \cdot$ & $Y_{11q_{1}}$ & $\cdot \cdot$ & $Y_{1j1}$ & $\cdot \cdot$ & $Y_{1jq_{j}}$ & $\cdot \cdot$ & $Y_{1n1}$  & $\cdot \cdot$ & $Y_{1nq_{n}}$ \\
			$	\vdots$ &$\vdots$	&  $\cdot \cdot$ &	$\vdots$ & $\cdot \cdot$ & 	\vdots& $\cdot \cdot$ & 	$\vdots$& $\cdot \cdot$ & $	\vdots$  & $\cdot \cdot$ & $\vdots$\\
			$ASV_{i}$ & $Y_{i11}$ &  $\cdot \cdot$ & $Y_{i1q_{1}}$ & $\cdot \cdot$ & $Y_{ij1}$ & $Y_{ijt}$   & $Y_{ijq_{j}}$ & $\cdot \cdot$ & $Y_{in1}$  & $\cdot \cdot$ & $Y_{inq_{n}}$\\
			$	\vdots$ &$\vdots$	&  $\cdot \cdot$ &	$\vdots$ & $\cdot \cdot$ & 	\vdots& $\cdot \cdot$ & 	$\vdots$& $\cdot \cdot$ & $	\vdots$  & $\cdot \cdot$ & $\vdots$\\
			$ASV_{m}$ & $Y_{m11}$ &  $\cdot \cdot$ & $Y_{m1q_{1}}$ & $\cdot \cdot$ & $Y_{mj1}$ & $\cdot \cdot$ & $Y_{mjq_{j}}$ & $\cdot \cdot$ & $Y_{mn1}$  & $\cdot \cdot$ & $Y_{mnq_{n}}$\\
		\end{tabular}}
	\end{table}

A count matrix ${\mY \in \mathbb{R}^{m \times N}}$ from a longitudinal microbiome study is displayed in Table \ref{tab01}. In this design, $Y_{ijt}$ denotes the abundance of an $i$-th ASV on the $j$-th 
subject at time point $t$. Moreover, $n$ is the number of subjects, ${q_{j}; j=1,2,\cdots,n}$ is the number of repeated observations from the $j$-th subject, $m$ is the total number of ASVs, $N=q_{1}+\cdots+q_{n}$ is the total number of biological samples.
	
\begin{table}
	\caption{\label{tab02}Sample data $\mX \in \mathbb{R}^{N \times p}$}
	\fbox{
		\begin{tabular}{c c c c c c}
			Sample ID & Subject ID & Time&$X_{1}$ & $\cdots$ & $X_{p}$\\
			\hline
			$s_{1,1}$ & $s_{1}$ & $t_{1}$&$x_{1,1}$ & $\cdots$  & $x_{1,p}$\\
			$	\vdots$ &$	\vdots$ &$	\vdots$ &$	\vdots$ &$	\vdots$&$	\vdots$\\
			$s_{n,q_{n}}$ & $s_{n}$ & $t_{q_{n}}$&$x_{N,1}$ & $\cdots$  & $x_{N,p}$\\
		\end{tabular}}
	\end{table}
	
Table~\ref{tab02} shows the associated sample information and $p$ covariates that are included in the matrix ${\mX \in \mathbb{R}^{N \times p}}$. To simplify the illustration, we assume that we are interested in the effect of a factor variable $X$ with two levels such as case and control. 

In order to identify a statistic ${\beta_{i}(\mY)}$, which measures the effect of $X$ on the abundance of $i$-th ASV, we consider the following marginal model for each ASV $i$: 	
\begin{eqnarray}
\label{eq:margmodel}
	\E(Y_{ijt}) & = & \mu_{ijt},\\
	\notag
	f(\mu_{ijt}) & = &x_{ijt} \beta_{i},
\end{eqnarray}
where $i=1,\cdots,m$, $j=1,\cdots,n$, $t=1,\cdots,q_{j}$, $\mu_{ijt}$ is the mean abundance, $\beta_{i}$ is the effect of explanatory variable $X$ on the average population abundance of $i$-th ASV, and $f(\cdot)$ is the variance-stabilizing transformation. Further, we assume that there is some order of dependence within each subject, so that 

\begin{equation}
\label{eq:cov-var}
	\text{cov}(Y_{ijk},Y_{ijl})  =
	\begin{cases}
	c_{i}(k-l) & ; k \neq l \hspace{.1in} \text{and} \hspace{.1in} j = 1, \cdots ,n\\
	\alpha_{i} & ;k = l \hspace{.1in} \text{and} \hspace{.1in} j = 1, \cdots ,n,
	\end{cases}
\end{equation}

where $c_{i}(.)$ is some covariance function. Note that \eqref{eq:margmodel} resembles the marginal model described in \cite{D2002}. 

To do the differential abundance analysis, first, we account for the unequal library size between samples using the median-ratio method by computing normalization constant $\delta_{ijt}$  \citep{AndersHuber2010}. Second, we fit a generalized linear model (GLM) type model with $arcsinh$ link function. When $f=arcsinh$ we can write model~\eqref{eq:margmodel} as:

\begin{eqnarray}
	\label{eq:margmodel2}
	\E(Y_{ijt})&=&\mu_{ijt},\\
	\notag
	\mu_{ijt}&=&\delta_{ijt} \cdot a_{ijt},\\
	\notag
	\log (a_{ijt}+(a_{ijt}^2+1)^{1/2})&=&x_{ijt} \beta_{i},
\end{eqnarray}

where $\log (a_{ijt}+(a_{ijt}^2+1)^{1/2})$ is the $arcsinh$ transformation of the expected abundance after accounting for library size.

Third, for each ASV, we estimate $\beta_{i}$ using the generalized estimating equation approach \citep{liang1986,halekoh2006} with a block bootstrap estimator of $c_{i}(k-l)$ \citep{lahiri2013}. Fourth, for the inferences on more than one ASVs, we obtain shrinkage estimators of $\beta_{i}$ by using the  empirical Bayes method \citep{Robbins1956,Stephens2016}, which is more stable than unshrunken  estimators \citep{E2012,L2014}. Now, we define the studentized version of $\hat{\beta}_{i}$:
 \begin{equation}
 \label{eq:pivot}
 {T_{i}=\frac{\hat{\beta}_{i}-\beta_{i}}{\text{SE}(\hat{\beta}_{i})}},
 \end{equation}
where $\text{SE}$ stands for standard error. 

To avoid the influence of nuisance parameters on the distribution of a statistic ($\hat{\beta}_{i}$), we use the studentized version of $\hat{\beta}_{i}$ that makes our test statistic approximately pivotal. Fifth, we approximate the sampling distribution of $\hat{\beta}_{i}$ and $T_{i}$ using the moving block bootstrap (MBB) method with a  pairwise nested double bootstrap.

We consider that each row in Table \ref{tab01} is $n$ independent time series of length $q_{j}; j=1,\cdots,n$. Let us assume that all the subjects have the same number of repeated observations $q_{j} = q$. For the MBB method, we draw block bootstrap realizations by preserving the dependency within-subject. Thus, for all the rows together in Table \ref{tab01}, we define the overlapping blocks of columns for each subject. 

For each subject, consider $\left(X , \mathbb{Y} \right),$ where $ \mathbb{Y}= \left\lbrace \bY_{1}, \cdots \bY_{q}\right\rbrace$ is the set of $q$ column vectors and $X$ be the covariate. We define the $v$-th block for each subject $ B_{v}  = \left\lbrace \bY_{v}, \cdots \bY_{v+l-1}\right\rbrace$ with the block size $l$, where $v=1, \cdots L$. The total number of overlapping blocks within a subject is $L=q-l+1,$ where $l$ is between one and $q $. Therefore, each subject is identified with $\mathbb{B} =\left\lbrace B_{1}, \cdots B_{L}\right\rbrace$ set of blocks of observations. We resample  $L_{0}\geq q/l$ numbers with replacement from $\left\lbrace 1, \cdots L \right\rbrace$ and obtain $\mathbb{B}^{*} = \left\lbrace B_{\tau(1)}^{*}, \cdots B_{\tau(L_{0})}^{*}\right\rbrace$, where $L_{0}$ is the smallest integer greater than $q/l$ and $\tau(1), \cdots \tau(L_{0})$ are resampled block numbers. We choose the first $q$ observations out of $L_{0} \cdot l $ to create a pairwise bootstrap realization $(X^{*},\mathbb{Y}^{*})$  with the same number of repeated observations $q$ as in $\left(X, \mathbb{Y}\right)$.

We repeat the blocking procedure with block size $l$ and pairwise resampling a large number of times,  obtaining $R$ bootstrap realizations. From this resampling procedure, we compute $R$ bootstrap estimates $\hat{\beta}^{(r)}_{i}; r=1,\cdots,R$ for each ASV. These provide the approximated sampling distribution of $\hat{\beta}_{i}; i=1,\cdots,m$. These estimates can be used to construct a confidence interval for $\beta_{i}; i=1,\cdots,m$.

As a refinement of the above method, the sampling distribution of $T_{i}$ is found by using the $R$ pairwise bootstrap samples and as many pairwise double bootstrap resamples for each pairwise  bootstrap realization $(X^{*},  \mathbb{Y}^{*})$. Thus, the standard error $\text{SE}_{*}(\hat{\beta}^{*}_{i})$ in $ T^{*}_{i} = \frac{\hat{\beta}_{i}^{*}-\hat{\beta}_{i}}{\text{SE}_{*}(\hat{\beta}_{i}^{*})} $ is estimated from the  pairwise  double MBB; $\text{SE}_{**}(\hat{\beta}^{**}_{i})$. Here $\text{SE}_{*}$ and $\text{SE}_{**}$ are conditional standard error on $(X,  \mathbb{Y})$ and $(X^{*},  \mathbb{Y}^{*})$, respectively. Finally, the approximated  sampling distribution of $T_{i}$ is used to compute the p-values. These p-values are then adjusted according to \cite{HB1995}'s algorithm that controls the false discovery rate (FDR). 

There is no analytical solution for the distribution of $T_{i}$,
so we check the invariance by simulation. In Figure \ref{fig01}, we show by a simulation that $T_{i}$ is approximately pivotal in the context of differential abundance analysis. 

Our next task is to determine the optimal block size for the given data, for which we propose a modified empirical subsampling procedure. 

\subsection{Choosing an Optimal Block Size}	

For each row $i$ in Table \ref{tab01}, let $\psi_{i}$ be the rejection probability of the bootstrap test under $\beta_{i}=0$ with the test statistic $T_{i}$ as our statistical functional. Because we are interested in a two-tailed test, we assume that $\psi_{i}$ is the two-sided probability $P\left(|T_{i}| \geq |k|\right)$, where $k$ is an observed value of $T_{i}$ based on the full dataset. Let $\hat{\psi}_{i}(q,l) = P_{*}\left(|T_{i}^{*}| \geq |k|\right)$  denote the estimate of $\psi_{i}$ using the moving block bootstrap (MBB) method on all repeated observations $q$ with block size $l$, where $P_{*}$ is the conditional probability given $\left(X, \mY \right)$ and $T^{*}_{i}$ is the bootstrap version of $T_{i}$.

Assume that we can create $W$ overlapping subsamples by choosing $\omega$ repeated observations from each subject. That is, $W = q-\omega +1$.  First, we will find the optimal block size for the subsample. Then using a modified version of the formula in \cite{HHJ1995}, we will compute the optimal block size for the full data.

We define the bias $\text{Bias}(\hat{\psi}_{i}(\omega,l))$ and the variance $\text{Var} (\hat{\psi}_{i}(\omega, l))$ and compute the mean squared error (MSE) of $\hat{\psi}_{i}(\omega, l)$ on the subsample. The optimal block size $l_{\omega}$ for each ASV in subsample is determined by balancing the trade-off between bias and variance of $\hat{\psi}_{i}(\omega,l)$. That is, 

\begin{equation}
\label{eq:i-th-MSEpsi}
l_{\omega} = \argmin_{l}\left\lbrace \text{MSE} (\hat{\psi}_{i}(\omega,l)): 1 < l < \omega \right\rbrace.
\end{equation}

In the differential abundance analysis, there are $m$ different ASVs. Thus, we define the optimal block size for the subsample such that $l_{\omega}$ minimizes the $l_{1}$-norm of \textbf{MSE} in estimating $\hat{\psi}_{i}(\omega,l)$. That is,

\begin{equation}
\label{eq:For-all-MSEpsi}
l_{\omega} = \argmin_{l}\left\lbrace ||\textbf{MSE} (\omega, l)||_{1}: 1 < l < \omega \right\rbrace, 
\end{equation}

where \[\textbf{MSE} (\omega,l)=\left[ \text{MSE}(\hat{\psi}_{1}(\omega, l)),\cdots \text{MSE}(\hat{\psi}_{m}(\omega, l))\right]^{T}.\]

Note that $l_{\omega}$ in \eqref{eq:For-all-MSEpsi} depends on the sampling distribution of 
\[
\hat{\bpsi}(\omega, l) = \left[\hat{\psi}_{1}(\omega, l),\cdots, \hat{\psi}_{m}(\omega, l)\right]^{T}.\]

Thus, we define a modified subsampling method to estimate $ \textbf{MSE}(\omega, l)$. We shall begin the procedure by choosing an initial block size $l_{I}$ for full data. We obtain $R$ bootstrap values of  $T^{(r)}_{i}(q, l_{I})$ on full data to compute   $\hat{\psi}_{i}(q,l_{I})$ for all ASVs by applying the MBB procedure on Table \ref{tab01}. 

Then, we let $\omega$ be the number of repeated observations from each subject to create $W$ subsamples of Table \ref{tab01}. Now, we choose a different block size than $l_{I}$, say, $l_{C} (< l_{I})$ to compute $R$ bootstrap values of $T^{(r)}_{i}(\omega, l_{C})$ by applying the MBB procedure on each subsample. From these bootstrap estimates, we compute  $\hat{\psi}_{i}^{(j)}(\omega,l_{C})$ for all ASVs on each subsample, where $j = 1, \cdots, W$. 

We can now estimate the MSE of $\hat{\psi}_{i}(\omega, l_{C})$ with $W=q-\omega+1$, the number of subsamples:    

\begin{equation}
\label{eq:MSE-one-row}
\text{MSE}\left(\hat{\psi}_{i}(\omega,l_{C})\right) = \dfrac{\sum\limits_{j=1}^{W} \left( \hat{\psi}_{i}\left(q,l_{I}\right)- \hat{\psi}^{(j)}_{i}\left(\omega,l_{C}\right)\right)^{2}}{W}.
\end{equation}

We compute \eqref{eq:MSE-one-row} for all $m$ rows in Table \ref{tab01}. Thus, we have 

\begin{equation}
\label{eq:MSE-all-rows}
	\textbf{MSE}\left(\omega,l_{C}\right)=\left[\text{MSE} \left(\hat{\psi}_{1}(\omega,l_{C})\right),\cdots, \text{MSE} \left(\hat{\psi}_{m}(\omega,l_{C})\right) \right]^{T}.
\end{equation}
	
We repeat the above procedure with different choices of $l_{C}$ on the subsamples. Then, we determine the optimal block size for the subsample, which minimizes the $l_{1}$-norm of $\textbf{MSE}\left(\omega,l_{C}\right)$:

\begin{equation}
\label{eq:opt-length-sub}
l_{\omega} = \argmin_{l_{C}} \left\lbrace ||\textbf{MSE} \left(\omega,l_{C}\right)||_{1} ; 1 < l_{C} < l_{I}\right\rbrace.
\end{equation}

Note that $l_{\omega} $ in~\eqref{eq:opt-length-sub} is the optimal block size for the subsample with $\omega$ repeated observations in each subject. Because MSE is proportional to the number of repeated observations, we scale up the optimal block size for the subsample until obtaining the optimal block size for the original data in Table \ref{tab01}:

\begin{equation}
\label{eq:opt-length-ori}
l_{o} = \left(\frac{q}{\omega}\right)^{1/\nu} \cdot l_{\omega},
\end{equation} 
where $\nu$ is five if $\psi$ is the two-sided probability  \citep{HHJ1995}.

It is necessary to make sure that MSE measures the uncertainty associated with different block sizes. In the differential abundance context, because the number of repeated observations per subject is different in practice, we choose a reasonable proportion $\omega$ of repeated observations from each subject to compute the MSE. Thus, the optimal block size is 

\begin{equation}
\label{eq:opt-length-not-eq-length}
l_{o}=\left(\frac{1}{\omega}\right)^{1/\nu}l_{\omega}.
\end{equation}

Based on the simulation study, we recommend choosing $\omega$ to compute MSE over at least five subsamples. 

In practice, we choose $l_{I}$ using a partial autocorrelation (PAC) plot. First, we use the variance-stabilizing transformation on the original abundance table. Then, for each ASV, we compute the average transformed abundance at each time point. Finally, we plot within-subject PAC for each ASV. We can choose the top six ASVs to plot based on the total transformed abundances. Using the PAC plots, we consider the first occurrence of a lag with sufficiently close to zero PAC ($< .25$) as initial block size $l_{I}$. There might be a spurious higher PAC at larger lags due to a small number of observations or abundances close to zero that can be diagnosed using lag-plots.

We can also choose $l_{I}$ using a lag-plot that identifies autocorrelation patterns in time series data. We extend this plot to visualize the within-subject autocorrelation in longitudinal microbiome data. For each ASV, we plot transformed abundances at different lags. The lag-plot is interpreted as follows: 1) lack of autocorrelation is implied by the absence of a pattern in the lag plot; 2) weak to moderate autocorrelation is implied by less clustering of points along the diagonal, and 3) high autocorrelation is implied by tight clustering of points along the diagonal.

	\section{Results and Discussion} \label{simulation}	
	In this section, we illustrate the moving block bootstrap (MBB) inference for longitudinal microbiome data with different orders of dependence using synthetic and real data.

\subsection{Synthetic Data}

We simulated longitudinal microbiome data at two levels (case/control) of a group factor variable, although the MBB method could be used for testing multiple levels. We started by estimating realistic parameters (mean and over-dispersion parameters) for the gamma-Poisson on real data from the study in \cite{DCM2015}. 

We simulated stationary abundances over time in two different settings with 50 (Setting-Z) and 100 (Setting-ZL) ASVs with 50 \% and 20 \% differentially abundant. In both settings, we simulated ASVs abundances from an autoregressive process (AR),  closely resembling microbiome data \citep{DCM2015, F2017, Proctor2018}. Within each setting, we used three different order of dependence: (i) AR(1), (ii) AR(2)  and (iii) AR(1) for control and AR(2) for case subjects\\
(i) $X_{n}= .8 X_{n-1} + Z_{n}$ (dep-order1) \\
(ii) $X_{n}=.5 X_{n-2}+ .3 X_{n-1} + Z_{n}$ (dep-order2) and \\
(iii) $X_{n}= .8 X_{n-1} + Z_{n}; X_{m}= .5 X_{m-2}+ .3 X_{m-1} + Z_{m}$ (dep-orders 1\&2).

The innovations were $Z_{n} \sim \text{Negative Binomial}$. In Setting-Z, we simulated 10 repeated observations from 10 subjects for each level (case / control), giving a total of  200 simulated observations. In Setting-ZL, we kept the order of dependence similar to Setting-Z but increased the number of ASVs, decreased the percentage of differentially abundant ASVs, and increased repeated observations to 15. 

Each such set-up was simulated 50 times and provided estimates of the true and false positive rates for all possible values of false discovery rate (FDR) cut-off values. Because we have two levels of the covariate, we use a receiver operating characteristic (ROC) curve to visualize the true positive rate (TPR) and false positive rate (FPR) for all possible FDR cut-off values \citep{metz1978}. In the ROC curve, the better performance method produces the curve that is higher and to the North-West corner. If two ROC curves cross each other, then the better performance is determined by taking into account the true number of differential abundant ASVs and using a practical FDR cut-off value such as .05. 

\begin{figure}
	\centering
	\begin{tabular}{c}
		\makebox{\includegraphics[width=\textwidth, height=.43\textheight]{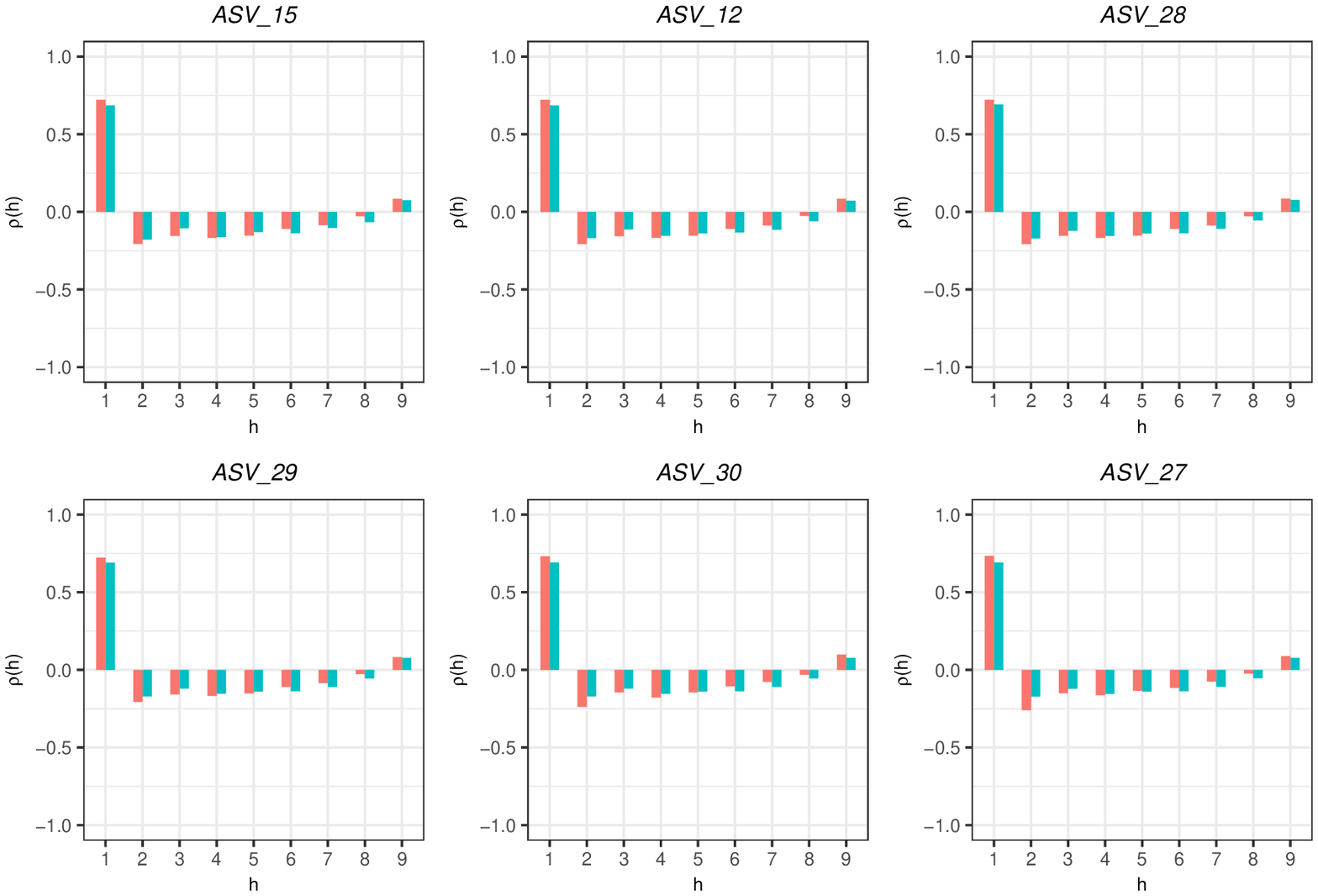}} \\
		\makebox{\includegraphics[width=\textwidth, height=.43\textheight]{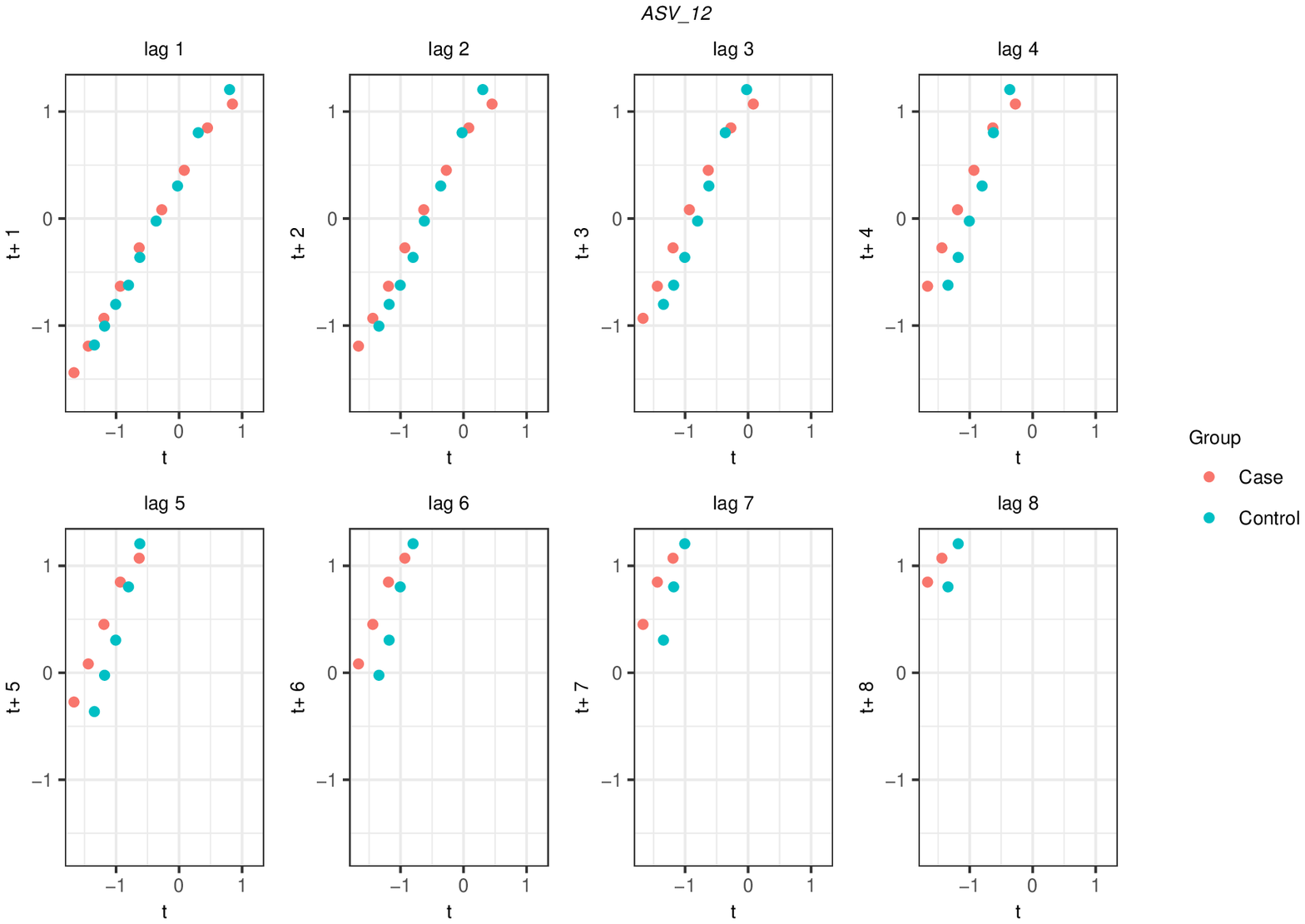} }  
	\end{tabular}
	\caption{ \label{fig02}\textbf{Top:} Each facet shows a PAC plot for different ASVs in Setting-Z (dep-order1) and colors refer to the level of a group variable (case/control). For the six top total arcsinh transformed abundance ASVs, PAC has a larger spike at lag-1, then decreasing after lag-4 for case (red) and control (green). We can choose initial block size $l_{I} = 5$ (lag four is equal to block size 5). We can look at the lag-plot for ASV\_12; \textbf{Bottom:} The lag-plot for ASV\_12 shows points make less clustering along the diagonal after lag-3. We can still choose $l_{I} = 5$.}
\end{figure}

In the MBB, we defined as our pivotal quantity the studentized shrinkage estimator and used the subsampling procedure to determine the optimal block size as explained in Section \ref{method}. We used the partial autocorrelation (PAC) plot and lag-plot at different lags to specify the initial block size. Figure \ref{fig02} shows an example (Setting-Z, dep-order1) of a PAC plot for six selected ASVs. In the top six ASVs based on the total of arcsinh transformed abundances, we observed that PAC  sufficiently approaches to zero at lag four, which suggesting a block size of five. We noticed a spurious larger PAC at larger lags due to a few observations that are visualized in the lag-plot of selected ASVs. We presented PAC and lag-plots for all the settings in the supplement. 

Because there are 10 repeated observations from each subject, we used 60\% of repeated observations from each subject to create five (10-6+1) overlapping subsamples. Then, we computed the average mean squared error of two-sided probability over these five subsamples for block sizes 2, 3, and 4. Table \ref{tab03} shows the frequency distribution for optimal block size for full data in each set-up in Setting-Z. We noticed the optimal block size distribution has smaller variability in each set-up and a single mode, except in Setting-Z(dep-orders 1\&2). 

\begin{table}
	\caption{\label{tab03} Frequency distribution for the optimal block size estimation for each set-up in Setting-Z.}
	\fbox{%
		\begin{tabular}{ l l l l l }
			\multirow{2}{*}{Set-up} & \multicolumn{3}{c}{Estimated block size}\\
			\cline{2-4}
			& 2 & 3 & 4  \\ 
			\hline
			(i) &.20 &\textbf{.42}  &.38   \\  
			(ii) & .34& \textbf{.38} &.28     \\ 
			(iii) & \textbf{.34}&.32  & \textbf{.34 }
		\end{tabular}}
	\end{table}

\begin{figure}
	\centering
	\begin{tabular}{cc}
		\hspace{-.5in}
			\makebox{\includegraphics[width=.65\textwidth, scale = 1.5]{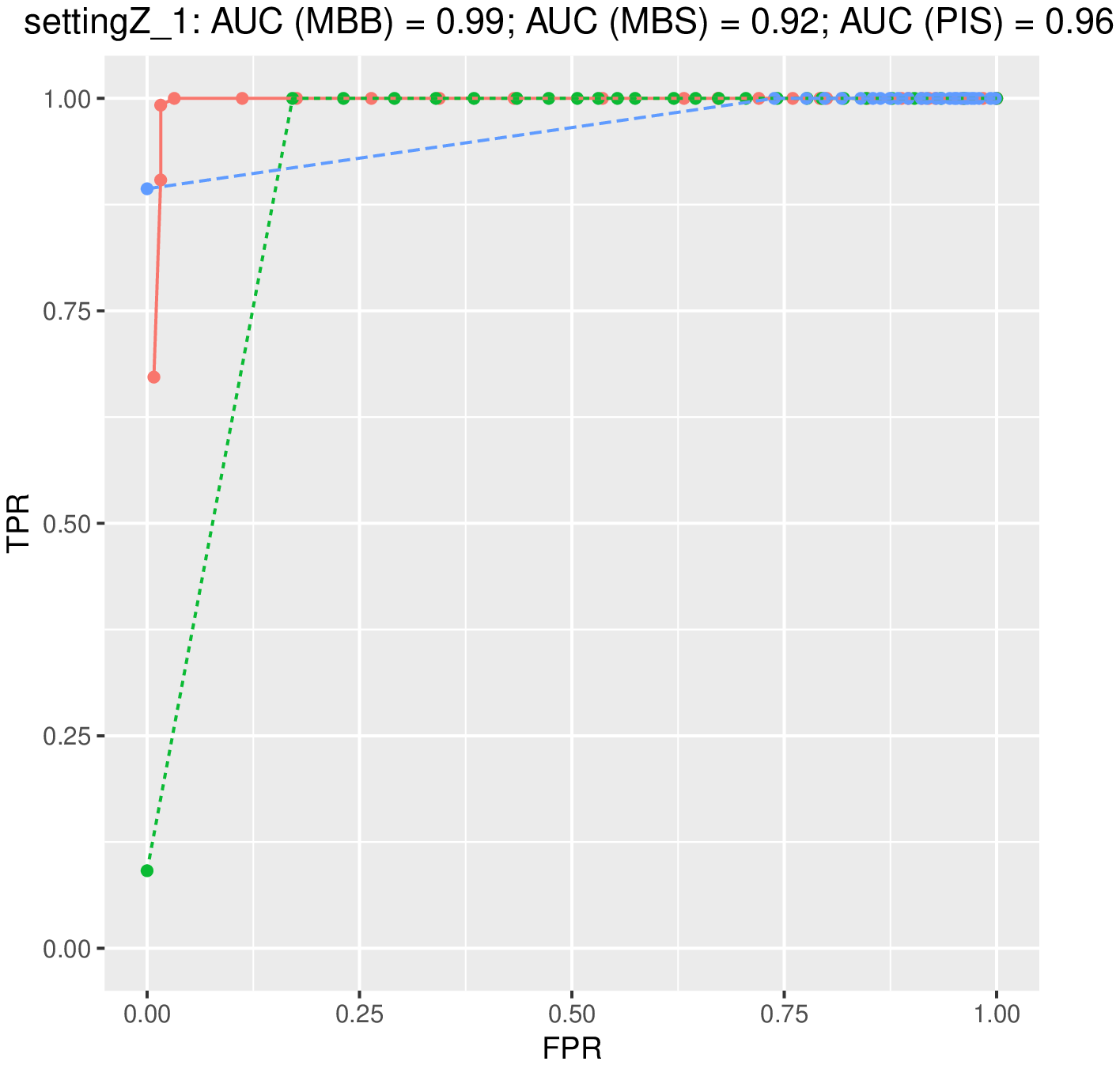}} &
		\hspace{-1.1in}
			\makebox{\includegraphics[width=.65\textwidth]{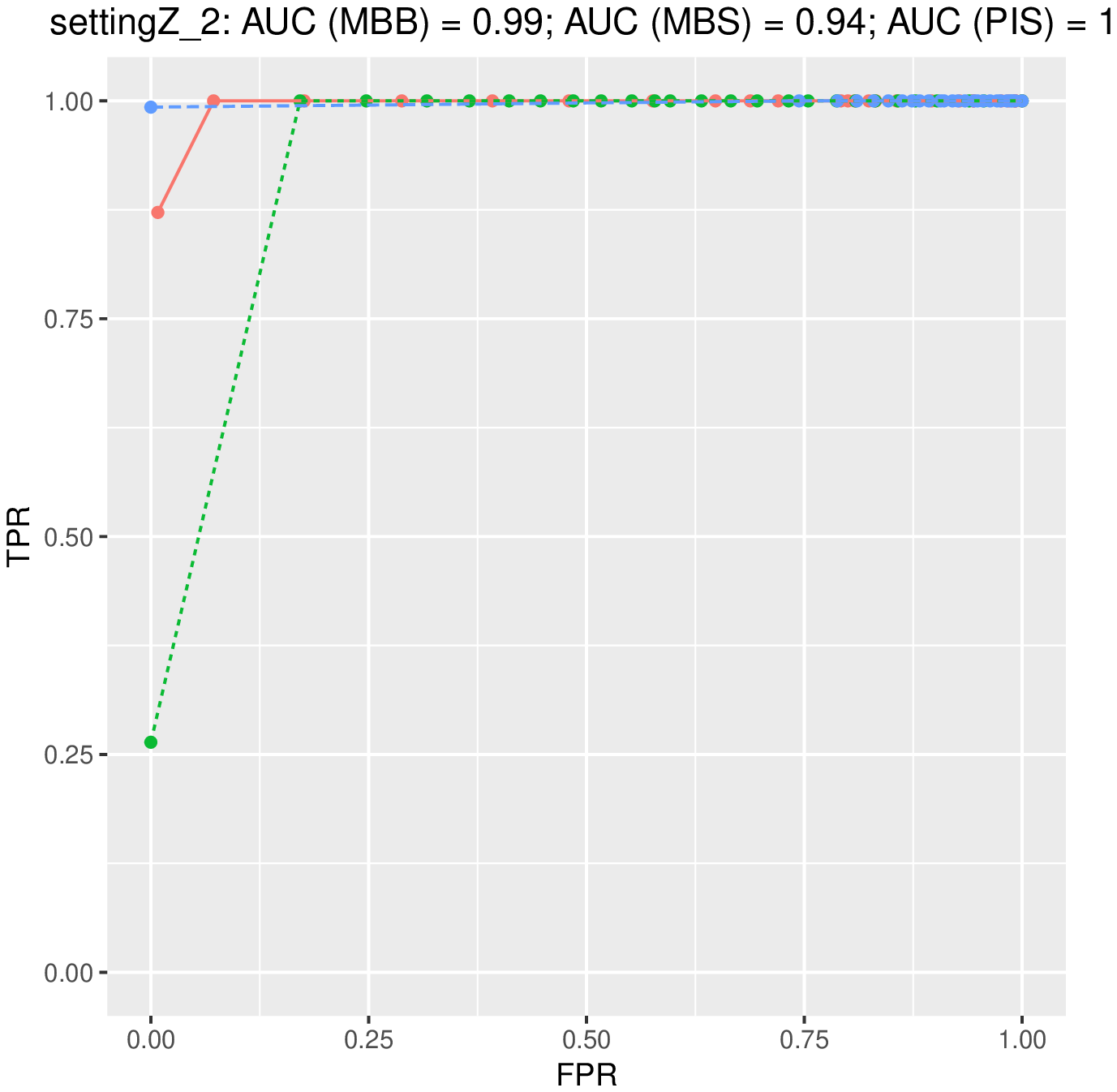} }  \\
		\multicolumn{2}{c}{	\makebox{\includegraphics[width=.65\textwidth, scale = 1.5]{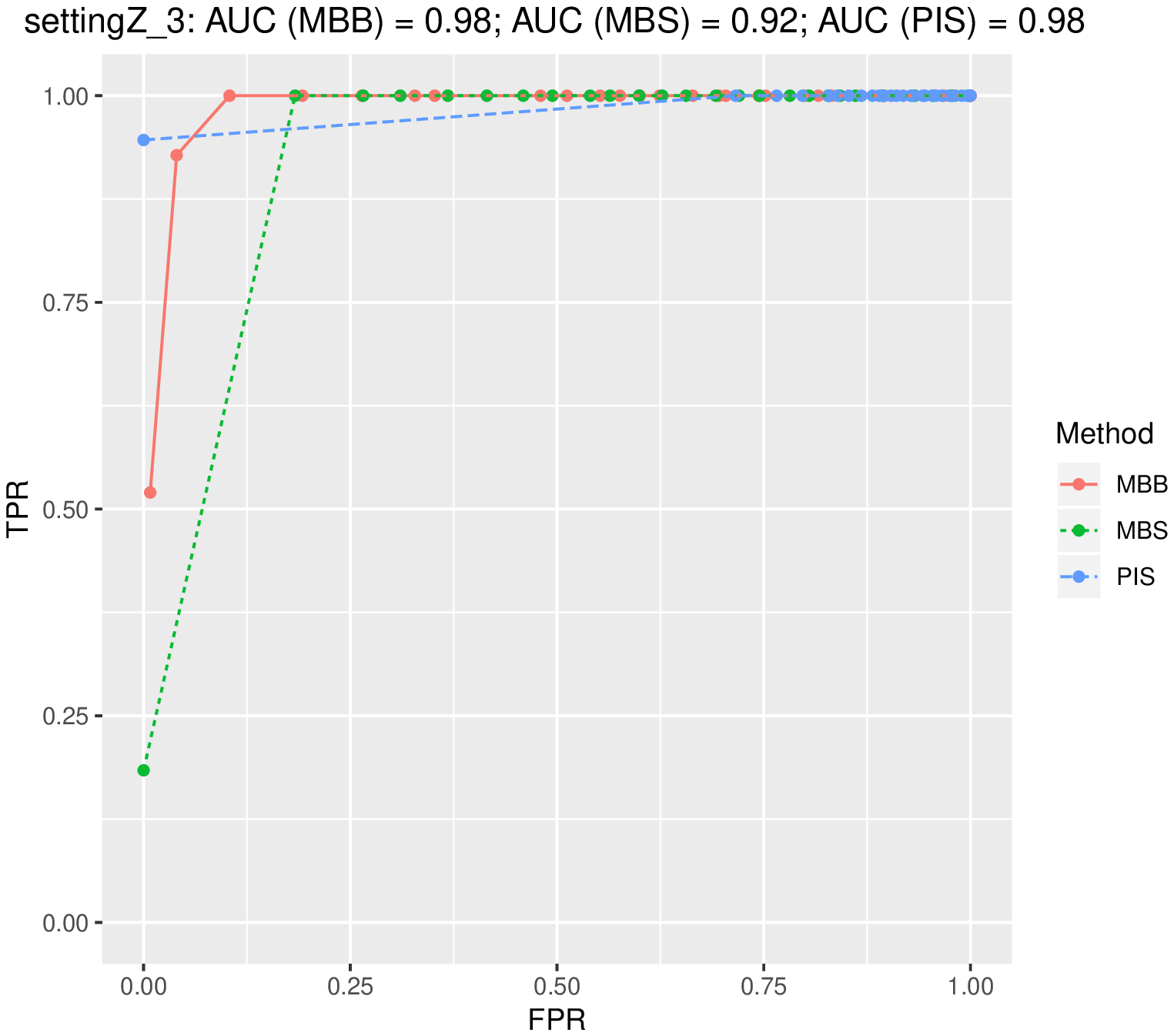}}}
	\end{tabular}
	\caption{ \label{fig03}ROC curve in Setting-Z (all three set-ups) for MBB (red), MBS (green), and PIS (blue) methods.}
\end{figure}

For each simulation run in Setting-Z, we ran the MBB method to compute the adjusted p-values with the optimal block size. Then, we computed the false positive rate (FPR) and true positive rates (TPR) based on the truth for all possible FDR cut-off values. Figure \ref{fig03} shows the ROC curve for Setting-Z in all set-ups.  In all three set-ups with 50\% true differential abundant ASVs, MBB works better than MBS that is the most conservative procedure. For an FDR cut-off of .05, that should be in the lower left of ROC curve, PIS produces more TPR than MBB. We conclude that overall MBB outperforms MBS and PIS in Setting-Z for practical FDR cut-off values.

\begin{figure}
	\centering
	\begin{tabular}{c}
			\makebox{\includegraphics[width=\textwidth, height=.43\textheight]{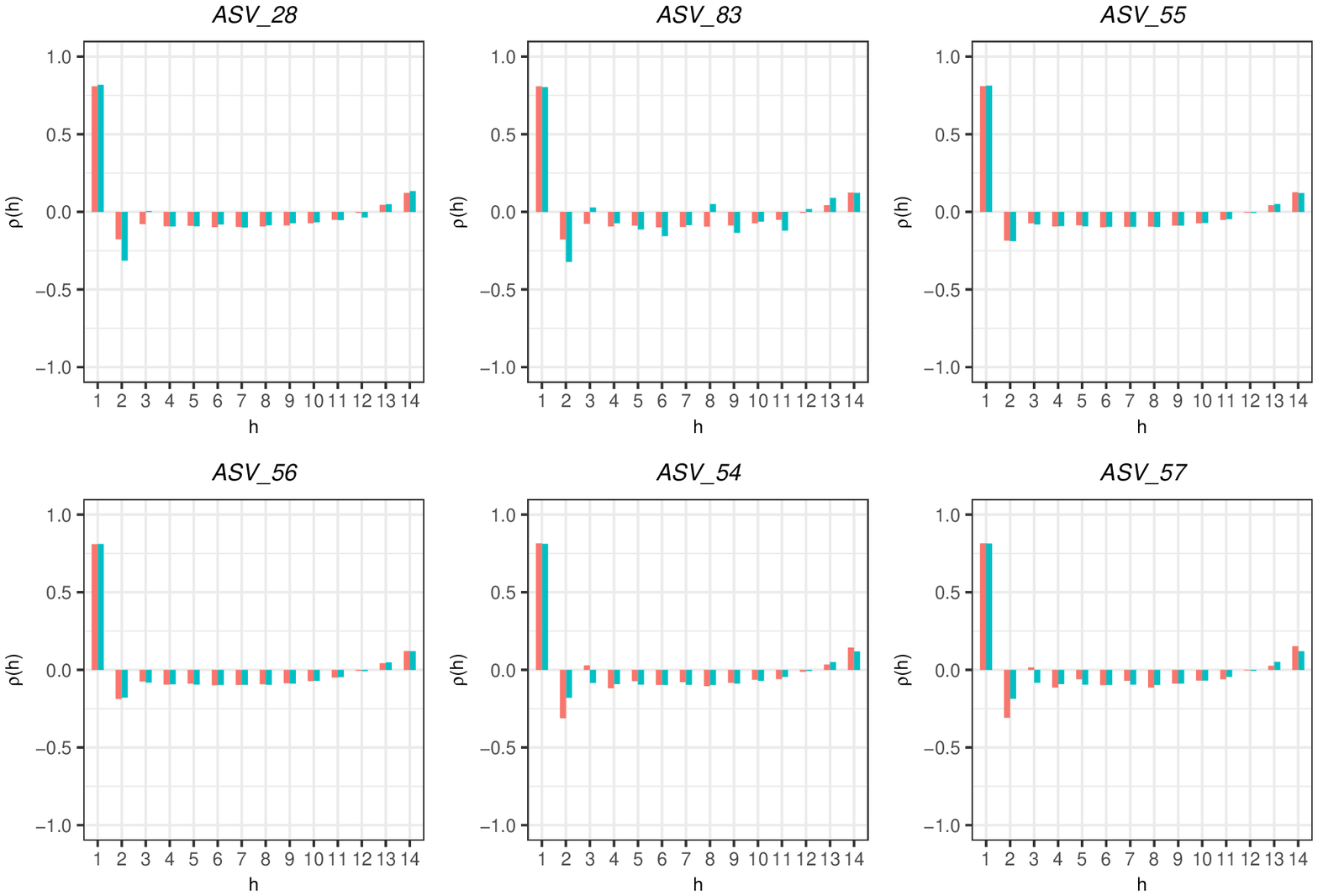}} \\
			\makebox{\includegraphics[width=\textwidth, height=.43\textheight]{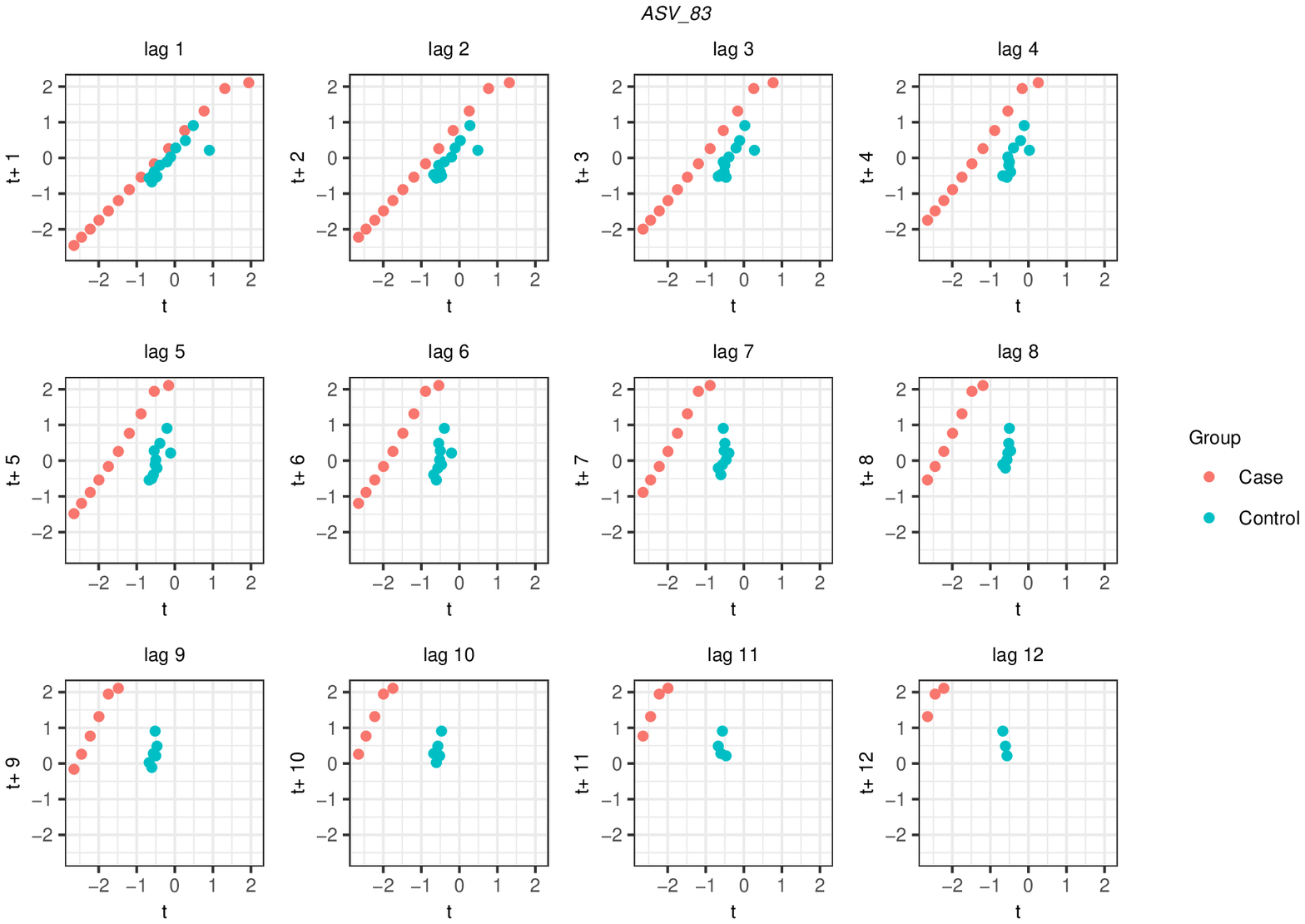} }  
	\end{tabular}
	\caption{ \label{fig04} \textbf{Top:} Each facet shows a PAC plot of different ASVs in Setting-ZL (dep-order1) and colors refer to the level of a group variable (case/control). For the ASV\_83, PAC has a larger spike at lag-1, then decreasing until lag-6 for control (green) and increasing at lag-6. We can choose initial block size $l_{I} = 7$. We can look at the lag-plot for ASV\_83 to check whether PAC at lag-6 is a spurious effect; \textbf{Bottom:} The lag-plot for ASV\_83 shows points make less clustering along the diagonal after lag-6 in control. We can still choose $l_{I} = 7$.}
\end{figure}

In Setting-ZL, we simulated data with 100 ASVs and of which only 20\% were differentially abundant and 15 repeated observations. For this Setting-ZL, we used the PAC plot and lag-plot to choose the initial block size of seven. An example is in Figure \ref{fig04}. We used 70\% of repeated observations for the subsampling procedure to make five (15-10+1 = 6) overlapping subsamples. 

\begin{table}
\caption{\label{tab04}Frequency distribution for the optimal block size estimation for each set-up in Setting-ZL.}
\fbox{%
	\begin{tabular}{ll l l l l l}
		\multirow{2}{*}{Set-up} & \multicolumn{5}{c|}{Estimated block size}\\
		\cline{2-6}
		& 2 & 3 & 4 & 5& 6 \\ 
		\hline
		(i) &.12  &.10  & \textbf{.28} &.22 &\textbf{ .28}\\  
		(ii) &.18 & .20 & .16 &.22 &\textbf{.24} \\ 
		(iii)  &.18  &.20 &.20 &.18& \textbf{.24}
	\end{tabular}}
\end{table}
	
Table \ref{tab04} shows the frequency distribution of optimal block size. This distribution has less variability, and the mode(s) in each set-up in Setting-ZL is larger than the corresponding set-up in Setting-Z. This supports the fact that the optimal block size increases as the number of repeated observations.

\begin{figure}
	\centering
	\begin{tabular}{ll}
		\hspace{-.5in}
		\makebox{\includegraphics[width=.65\textwidth, scale=1.5]{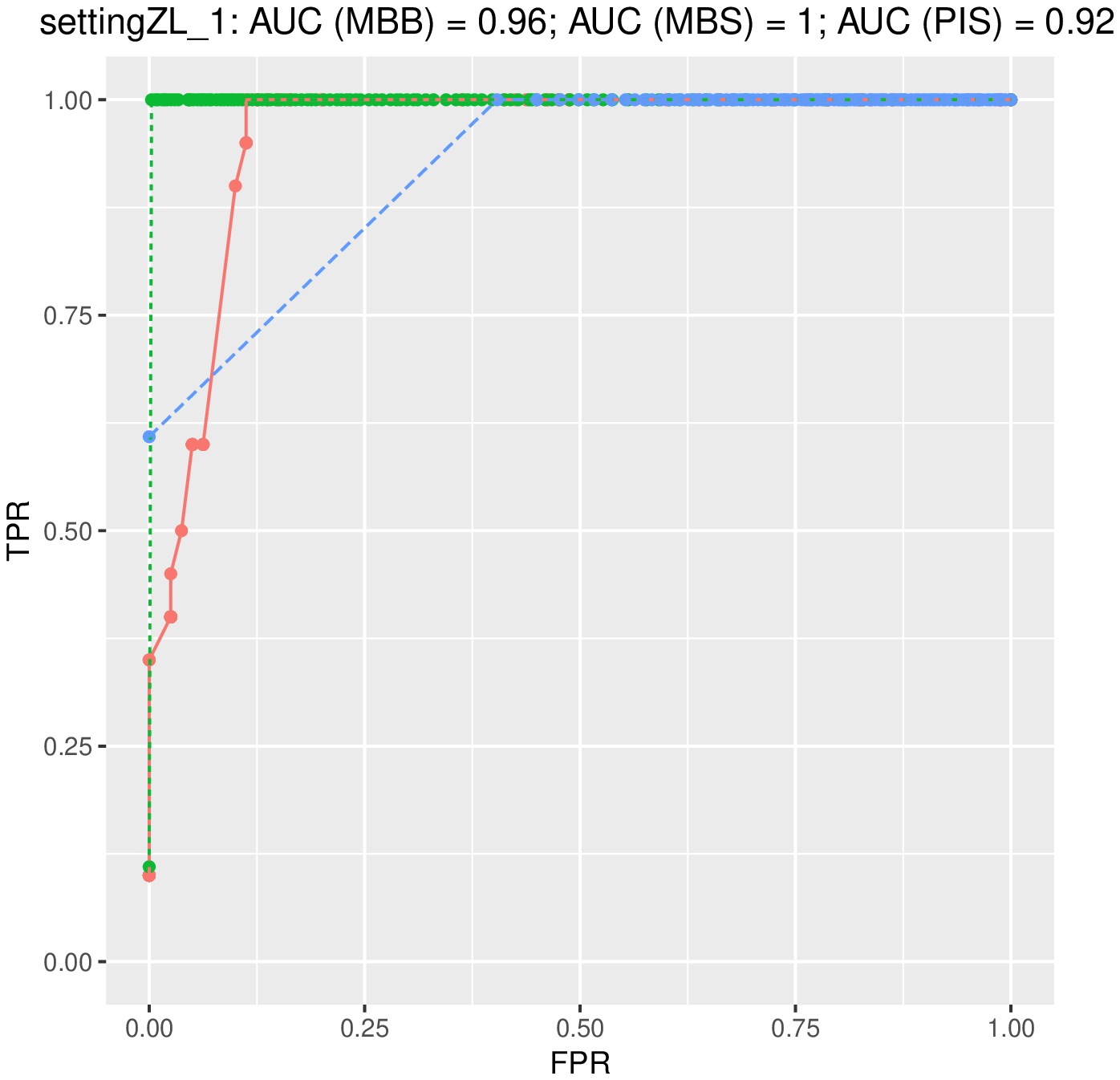}} &
		\hspace{-1in}
		\makebox{\includegraphics[width=.65\textwidth, scale = 1.5]{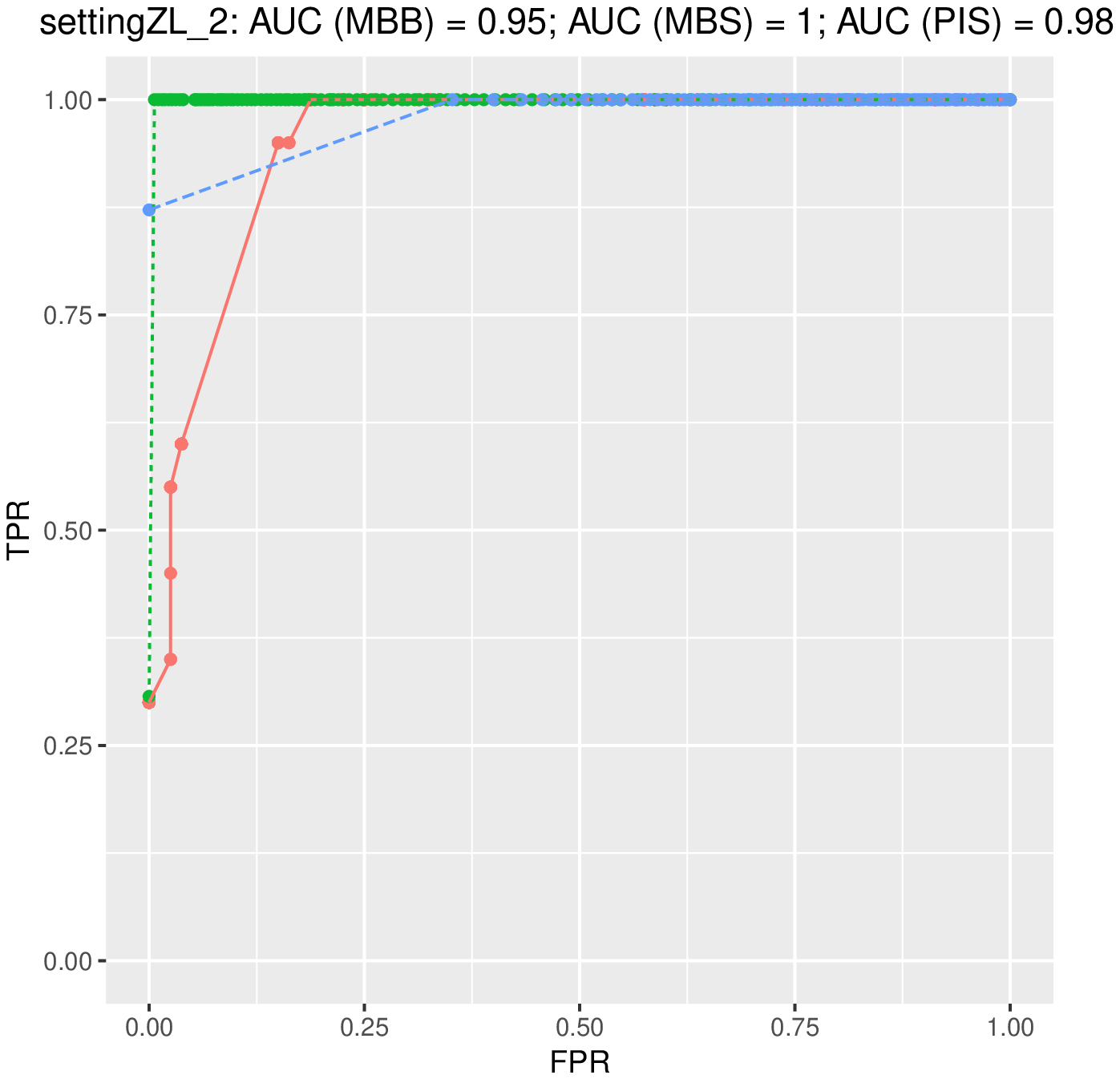} }  \\
		\multicolumn{2}{c}{
			\makebox{\includegraphics[width=.65\textwidth, scale = 1.5]{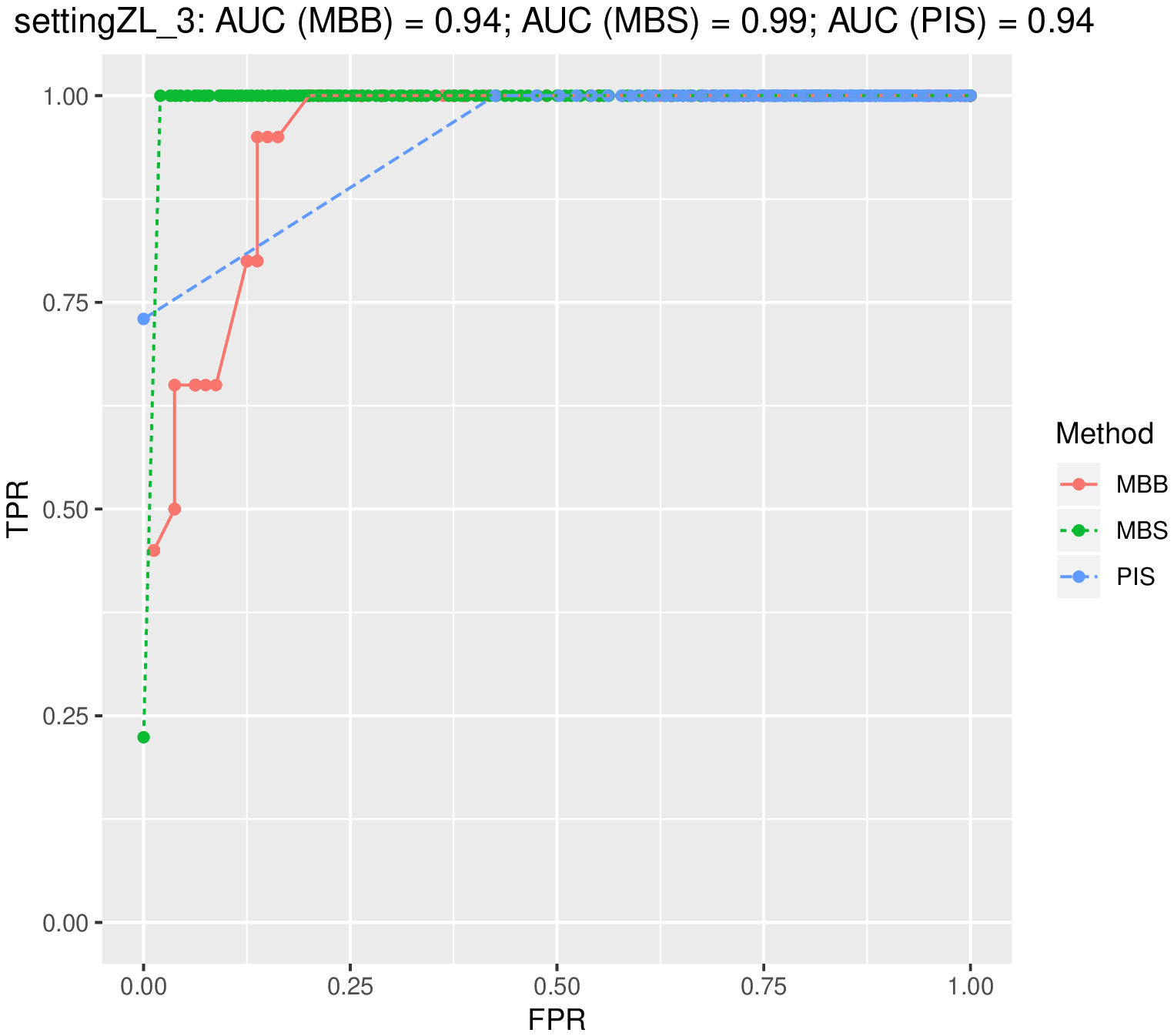}}}
	\end{tabular}
	\caption{ \label{fig05}ROC curve in Setting-ZL (all three set-ups) for MBB (red), MBS (green), and PIS (blue) methods. }
\end{figure}

In Figure \ref{fig05}, we show the ROC curve comparison for Setting-ZL in all set-ups. The MBS method performs better than MBB and PIS when there is less percentage of true differential abundant ASVs. However, in the bottom figure, the MBS method has smaller TPR than MBB at lower FDR cut-off values. Moreover, PIS has higher FPR than MBB for small FDR cut-off values (.05). For example, in the top left ROC curve, (FPR, TPR) = (0, .65) for PIS when FDR = 0, but (FPR, TPR) = (.35, 1) when FDR  = .01. To avoid the situation where almost all significantly differential abundant ASVs are false positives, we must keep FPR small for practical FDR cut-off values.

We concluded that the MBB method has high TPR and small FPR for practical FDR cut-off values in both settings.

\subsection{Differential Abundance Analysis of Stanford Pregnancy Data-A}

The Stanford pregnancy data \citep[hereafter, Stanford-A]{DCM2015} were the key motivation for the development of this method. The cohort consists of 761 biological vaginal samples collected from 40 pregnant women over gestation. These data included 33 term and 7 preterm subjects. We considered the differential abundance analysis of preterm versus term birth. Thus, the number of covariate was $p = 1$.

\cite{DCM2015} discussed a differential abundance analysis based on a merge-by-subject (MBS) approach and mentioned the loss in power compared to a potential analysis which would account for dependency within each subject. \cite{DCM2015} defined the community state type 4 (CST4) samples and showed that only ASVs from \textit{Gardnerella} and \textit{Anaerococcus} genera were significantly abundant by MBS analysis.  

We reanalyzed the Stanford-A data without subsetting CST 4 samples. We identified 598 ASVs in all samples. We excluded marginal preterm subjects, which were defined as women undergoing labor during the 37th gestational week. This process resulted in 678 samples, including 28 term and 7 preterm subjects. Then, we chose 109 ASVs that were present in at least 5\% of the samples. We tested each ASV to identify differentially abundant ASVs.  We observed more than one strain with the same name in Table \ref{tab05} and \ref{tab06} and used the strain identification number to distinguish among these variants.

Table \ref{tab05} shows differentially abundant ASVs from the MBS method. We observed that increased \textit{Gardnerella strain 22} and \textit{Prevotella strain 14} and decreased \textit{Lactobacillus gasseri strain 53} abundances were risk factors for preterm birth at FDR .05 in the Stanford-A data.  

We used the MBB method to identify the differentially abundant ASVs in the Stanford-A data. First, we identified the initial block size as nine using the PAC plot in Figure \ref{fig09}. We used 70\% of repeated observations within each subject to determine the optimal block size. The mean squared error for each block size $2,\cdots,8$ was calculated over 12 overlapping subsamples. Finally, with the optimal block size five as in Figure \ref{fig12}, we used $R=200$ MBB realizations and $RR=50$ double MBB realizations to find differentially abundant ASVs at FDR cut-off of .05.

In Figure \ref{fig20}  we visualized the arcsinh transformed abundances of significant ASVs that have $|\hat{\beta}_{i}| >1$ using the MBB method. Among these ASVs, \textit{Megasphaera strain 104}, \textit{F:Dethiosulfovibrionaceae strain 46}, \textit{Veillonella strain 106}, \textit{Clostridium strain 86}, \textit{Sneathia strain 44}
seemed significant due to technical error or other sampling artifact because these ASVs were present in few preterm or term subjects.

Table \ref{tab06} shows that at FDR .05, subjects with subsequent preterm labor in Stanford-A cohort were associated with significantly increased abundances in bacterial vaginosis-related (BV) ASVs and decreased \textit{Lactobacillus} species abundances. The large number of ASVs with increased abundances in the subjects and $|\hat{\beta}_{i}| >1$ with preterm labor that were identified with the MBB method makes sense, given the suspected high diversity of the risk-associated \textit{Gardnerella}-rich, \textit{Lactobacillus}-poor vaginal communities and their assumed ecological interactions.

\subsection{Differential Abundance Analysis of Stanford Pregnancy Data-B }
	
Next, we considered the follow-up Stanford pregnancy data \citep[hereafter, Stanford-B]{CDGSCJBWDS2017} which were generated to test the findings in \cite{DCM2015} in a similar population. This consisted of vaginal samples from 29 term and nine preterm birth subjects. These 897 vaginal samples were collected at each gestational week. We used the higher resolution amplicon sequence variants (ASVs) identification method DADA2 \citep{CMRHJH2016} and identified 1537 ASVs. For the differential abundance analysis of preterm versus term birth, we filtered ASVs that were present in at least 10\% of 897 samples, resulting in 97 ASVs.

Table \ref{tab07} shows the differentially abundant ASVs at an FDR cut-off of .05 in the Stanford-B cohort using MBS analysis and noted that two different strains of decreased \textit{Lactobacillus crispatus} and \textit{Lactobacillus jensenii strain 5} abundances were risk factors for preterm.  

For the MBB procedure, we used the PAC plot in Figure \ref{fig10} to determine the initial block size as nine. Then, with the final optimal block size two as in Figure \ref{fig14}, we identified 25 differentially abundant ASVs as shown in Table \ref{tab08} with $|\hat{\beta}_{i}| > 1$.

In Figure \ref{fig21} we visualized the arcsinh transformed abundances of significant ASVs using the MBB method with $|\hat{\beta}_{i}| > 1$. We observed that \textit{Lactobacillus crispatus strain 8} was only present in one preterm subject at one gestational week, and its abundance over the gestational week was completely different from \textit{Lactobacillus crispatus strain 2}. Three strains from \textit{Peptoniphilus} were having similar abundances over the gestational week. \textit{Gardnerella strain 7} is a significant strain with larger abundance in preterm subjects but it was not identified by the MBS method.

Table \ref{tab08} shows that the risk factors of preterm birth are increased abundances of bacterial vaginosis-related ASVs and decreased \textit{Lactobacillus} species. Moreover, we observed that the increased abundance of \textit{Gardnerella} is a risk factor in both Stanford-A and B cohorts. Tables \ref{tab06} and \ref{tab08} show that the diversity of ASVs that are enriched in preterm subjects is greater than those ASVs that are depleted in preterm subjects using the MBB method.

\subsection{Differential Abundance Analysis of UAB Pregnancy Data}

We analyzed the  University of Alabama (UAB) pregnant women cohort to identify the ASVs related to the risk of preterm birth (PTB) in this different population. The Stanford-B and UAB cohorts have different racial profiles \citep{CDGSCJBWDS2017} and different risk for PTB. The UAB population had a prior history of PTB and received prenatal care. Using the DADA2 pipeline, we identified 2316 ASVs with the prevalence of at least one read. This cohort consisted of 1282 biological samples collected from 96 pregnant women (41 preterm and 55 term) weekly during gestation. We filtered ASVs that were present in at least 10\% of total biological samples. This filtering reduced the number of ASVs to 183. We considered the differential abundance analysis of preterm versus term birth. Thus, the number of covariate was $p = 1$. Table \ref{tab09} shows that the reduced \textit{Lactobacillus gasseri} from \textit{Lactobacillus} -vaginal community and diverse bacterial vaginosis-related species are differentially abundant using the MBS method. 

We used the MBB method to analyze the UAB cohort data. We used the PAC plot in Figure \ref{fig11} to identify the initial block size as eleven. We used 80\% of repeated observations within each subject to create six overlapping subsamples for subsampling method. We chose a larger percentage than the last two applications because there was a larger number of repeated observation from each subject than the other two applications. With the optimal block size eight as in Figure \ref{fig13}, we identified 45 differentially abundant ASVs at FDR .05. However, among these ASVs, only 13 ASVs have larger than one absolute log-fold change in arcsinh scale.

Table \ref{tab10} shows that in contrast to the MBS analysis, the risk of preterm is significantly associated with increased abundances of bacterial vaginosis-related ASVs but no \textit{Gardnerella} in the UAB cohort with $\hat{\beta} > 1$. Further, the reduced \textit{Lactobacillus} species with $|\hat{\beta}| > 1$ were not identified as risk factors for preterm subjects in the UAB cohort. Some possible reasons for these findings are that microbiome composition in this cohort might be highly disturbed by prenatal care or had the different normal vaginal microbiome composition than the Stanford cohorts.

\begin{figure}
	\centering
	\includegraphics[width=\textwidth]{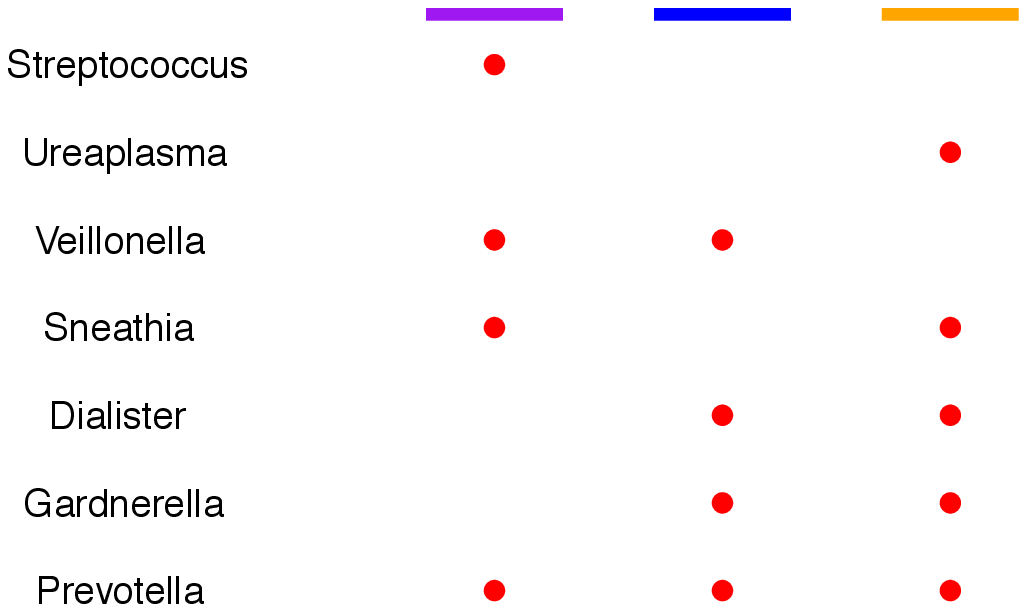}%
	\vfill
	\includegraphics[width=\textwidth]{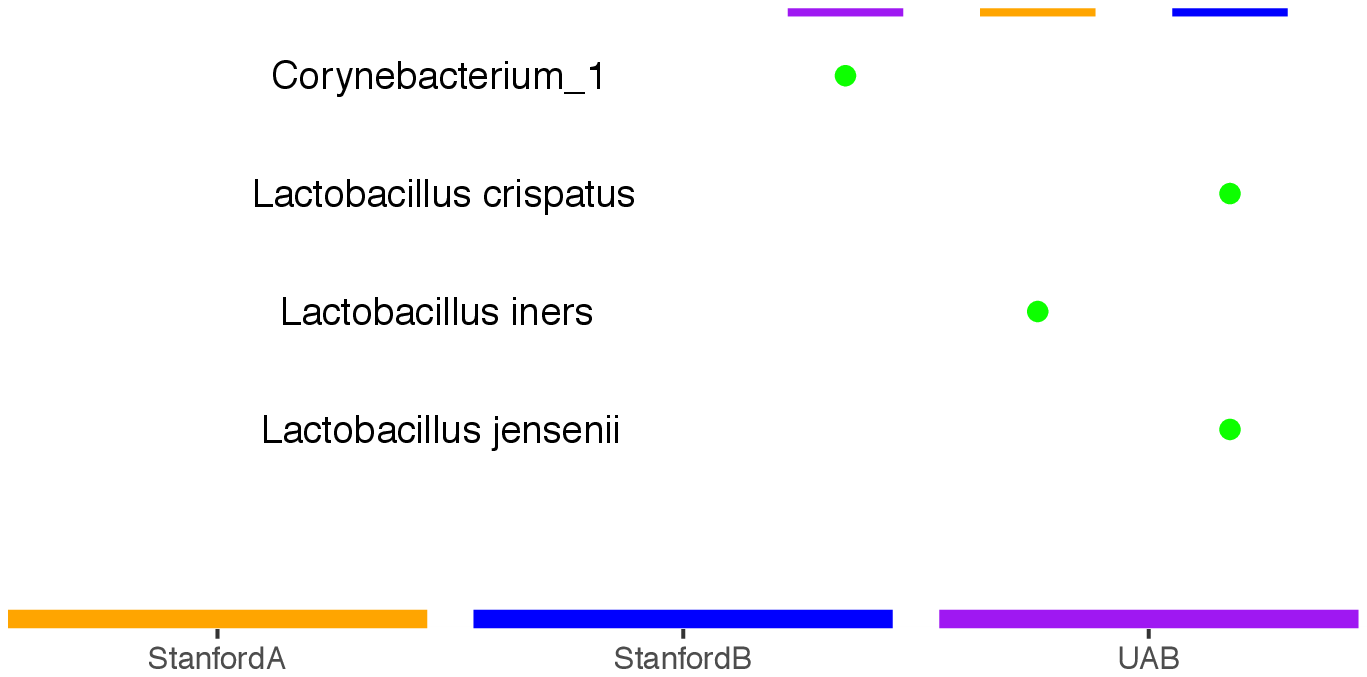}
	\caption{\label{fig06}Comparison of eight differentially abundant genera and \text{Lactobacillus} species across Stanford-A, Stanford-B and UAB cohorts at FDR .05 that are associated with bacterial vaginosis. Red color refers to significantly increased and $\hat{\beta}>1$ abundances and green color refers to decreased and $\hat{\beta} < - 1$ abundances of genera and \textit{Lactobacillus} species in preterm than term subjects. For example, increased abundances (red) of the \textit{Gardnerella} genus is a risk factor in Stanford-A and Stanford-B cohorts but not in UAB cohort. }
\end{figure}

In Figure \ref{fig06}, we compare results from the Stanford-A, Stanford-B, and UAB cohorts. We visualized some bacterial vaginosis-related genera, \textit{Lactobacillus} species, and the most differentially abundant \textit{Cornyebacterium} genus in the UAB cohort. The intersection matrix in Figure \ref{fig06} shows that increased abundances of \textit{Prevotella} associated with preterm labor are the risk in all three cohorts while increased \textit{Gardnerella} and \textit{Dialister} genera abundances are common risk factors in Stanford cohorts. We also noted that decreased different \textit{Lactobacillus} species abundances are risk factors of preterm labor in Stanford cohorts but not in UAB cohort. These findings suggest that the microbial biomarkers for risk of preterm labor are increased bacterial vaginosis-related species and decreased \textit{Lactobacillus} species that depends on the racial profile. In addition, diversity of genera is increased in preterm subjects in Stanford cohorts.

\subsection{Differential Abundance Analysis of Human Oral Cavity}

Next, we did the differential abundance analysis of oral microbial communities of eight dentally healthy individuals from whom samples of the molars and incisors were collected on a daily basis over 29-31 consecutive days (Figure \ref{fig15}). Previous analysis of these data suggested not only that communities inhabiting the molars and incisors differ from one another but also that they tend to be relatively stable over the course of a month as assayed by the RV coefficient, a multivariate generalization of the correlation coefficient \citep{Proctor2018}. 

Here, we sought to determine the individual ASVs that differ in abundance by tooth class (molar vs. incisor) while exhibiting similar or different patterns of temporal variation. Thus, the covariate is tooth class $(p = 1)$. Since the initial analysis of these data indicated that communities differ on opposing aspects (buccal, lingual) of individual teeth we analyzed only communities on the lingual, or tongue-facing, tooth aspect. Similarly, the initial analysis indicated communities vary based on the jaw (upper, lower) so we restricted the present analysis to molars and incisors in the upper jaw. ASVs were filtered to retain only ASVs that were present in at least 25\% of total biological samples to reduce the effect of bias due to sampling. We took the mean abundances of 90 ASVs on the first molars (teeth 3,14) and the central incisors (teeth 8, 9) to represent molar and incisor communities, respectively. 

The MBS method identified 30 ASVs that were found to be differentially abundant between the molars and incisors at FDR .05 (Table \ref{tab11}). Of the 12 ASVs most strongly found to be enriched on the molars compared to the incisors 8 represented different strains of \textit{Prevotella}. In contrast, no \textit{Prevotella} strain was found to be over-represented on the incisors compared to the molars; rather, a variety of genera were represented (e.g., \textit{Actinomyces}, \textit{Capnocytophaga}, \textit{Corynebacterium}, \textit{Neisseria}, \textit{Rothia}, \textit{Streptococcus} and \textit{Abiotrophia}) with any given genus represented by at most 3 strains. 

Examining the distribution of individual ASVs over time revealed 4 notable patterns (Figure \ref{fig19}). First, several ASVs found by the MBS method were present in high abundance at the molars and incisors, including \textit{Corynebacterium durum strain 16}, \textit{Rothia dentocariosa strain 1}, \textit{Rothia strain 11}, and \textit{Actinomyces strain 10}. Second, some ASVs including \textit{Prevotella pallens strain 70} and \textit{Prevotella strain 76} appeared to spike and crash in abundance regardless of whether they were more abundant on the molars or incisors. Third, other ASVs appeared to be stable on the molars but experience temporal fluctuation on the incisors, such as \textit{Campylobacter concisus strain 55}, \textit{Prevotella nigrescens strain 49} and \textit{Prevotella melaninogenica strain 37}. Finally, rarely did differences between individuals appear to be the predominant factor driving the differences between the distributions. One exception includes \textit{Streptococcus sanguinis strain 5} which appeared to fluctuate around a stable mean for both molars and incisors in most individuals with the exception of Subject 1-107 who experienced periodic extinctions at the incisors. 
Another exception includes \textit{Abiotrophia defectiva strain 9} which appeared to crash to zero abundance at several time points on the incisors in subject 1-104. In these cases, the temporal variability may be a result of sampling or other technical artifacts rather than biological variability, though it is not possible for us to test this hypothesis explicitly.

\cite{utter2016} showed that individual ASVs in dental plaque vary between subjects and over time. Thus, we used the MBB method to improve the inference by accounting the underlying temporal variability within subjects. We used the PAC plot to identify the initial block size as eleven (Figure \ref{fig16}) and the optimal block size as four (Figure \ref{fig17}). The MBB method identified 52 ASVs that were differentially abundant when comparing molars and incisors, including 26 ($\hat{\beta}_{i} > 1$) that were over-represented on the molars and 11 that were over-represented on the incisors (Table \ref{tab12}). In addition to identifying the 8 Prevotella species that were more abundant on the molars, the MBB identified an additional 3 Prevotella species that were over-represented on the molars. Likewise, the MBB method identified not just one Veillonella strain, as the MBS method did, but 4 strains that were over-represented on the molars. In general, the MBB method identified a higher number of strains that distinguished between the molars and incisors. These observations are consistent with prior reports of high variability and inter-individuality of the oral microbiome in healthy individuals, which appeared to be mediated by strain level diversity \citep{utter2016}.

To gain an intuitive sense of the difference between the MBS and MBB method we plotted ASVs transformed abundance over time by subject for all ASVs identified as significant by just the MBB method (Figure \ref{fig18}). Several ASVs the MBB method detected were generally low abundance ASVs, including \textit{Prevotella oris strain 87}, \textit{Mogibacterium strain 72}, \textit{Solobacterium moorei strain 68}, and \textit{Peptostreptococcus stomatis strain 59}. Most transformed abundances appear to be relatively stable with the exception of the incisors or molars in certain individuals. For example, \textit{Veillonella strain 75} appears to fluctuate around a stable mean for both molars and incisors but is not observed at the incisors of 2 individuals (P1-9 and 1-104) at one or more time points. Other ASVs that follow similar population crashes include  \textit{Streptococcus strain 6} and \textit{Haemophilus strain 35}. It is unclear, at this point, whether these ASVs are not observed in these samples due to technical artifact or whether population-level extinctions occur in the healthy human oral cavity. 

One technical artifact we can test is whether the observed ASVs defined as different \textit{Veillonella} strains arise from independent organisms. \textit{Veillonella} species have 4 rRNA operons exhibiting notable sequence heterogeneity \citep{marchandin2003}, which may give rise to ASVs that are classified as different organisms merely due to the technical artifact. In such cases, the different ASVs would be autocorrelated in their distributions across time within sites. When analyzing the distributions of the distinct \textit{Veillonella} species strains 7, 21, 28 and 75 it is possible to see that they differ from one another by subject (Figure \ref{fig18}), suggesting that the sequence variations indeed arose from independent organisms. Thus, the different ASVs assigned to \textit{Veillonella} are unlikely to be spurious. Whether other technical artifacts may give rise to the observed patterns of temporal fluctuation in a select subset of individuals is unclear. A carefully designed follow up study should be undertaken to evaluate the source of these temporal fluctuations.

	\section{Conclusion} \label{con}
	Longitudinal microbiome data are used to either model abundance over time or compare the abundances of bacteria between two or more cohorts. We have devised a method for making nonparametric inferences in longitudinal microbiome data in the latter case. With the optimal block size computed using subsampling, moving block bootstrap (MBB) resamples with replacement the overlapping blocks within each subject to make bootstrap realizations. Then, the MBB method computes the sampling distribution of the chosen pivotal quantity to draw valid inferences. Finally, it ranks bacteria for follow-up clinical studies based on the adjusted p-values.

We use exploratory tools to set-up the two tuning parameters: (i) initial block size and (ii) number of repeated observations for subsampling. These two tuning parameters may limit the MBB method to use for longitudinal design with at least five repeated observations. In addition, temporal variability in microbiome data may consist of technical and biological variabilities. Thus, we use appropriate pre-filtering to remove the unwanted noise. For example, we used 5\% , 10\% , 10\% and 25\% pre-filtering for Stanford-A, Stanford-B, UAB, and oral data, respectively.

 Although our method is computationally intensive, an accurate inference can be executed using parallel computing in R which as implemented in the open-source R package \href{https://github.com/PratheepaJ/bootLong}{https://github.com/PratheepaJ/bootLong}. 

Compared with parametric model-based approaches, MBB has flexibility in handling heterogeneity and temporal dependence structure, whereas parametric methods have to define a complete model dependency. In addition, compared with MBS (merge-by-subject) and PIS (presume independent samples) methods, the advantages of the MBB method are high true positive rates and small false positive rates for practical FDR cut-off values. The proposed method can be easily extended to make inferences on many genomic settings (RNA-seq, single cell studies as well as metagenomic data analyses) making biomedical research more reproducible.

	\section*{Acknowledgments}
	This work was partially supported by the NIH (R01 grant AI112401 to D.A.R., S.P.H.), the NSF (DMS 1501767 to S.P.H.), and the March of Dimes Prematurity Research Center  (B.J.C. and P.J.).

	\bibliographystyle{agsm}
	\bibliography{ms}
	\newpage
	\section*{Supplemental Material}
	\beginsupplement
\subsection*{Statistical Terms:}
\begin{itemize}
	\item Pivotal statistic: a pivotal statistic is a function of the observed data and the parameter of interest that has a distribution that does not depend on unknown parameters under the null hypothesis \citep{CH1979}. For example, the t-statistic is a pivotal quantity when the parameter of interest is the population mean. 
	
	\begin{figure}
		\centering
		\makebox{\includegraphics[width=\textwidth, scale = 1.5]{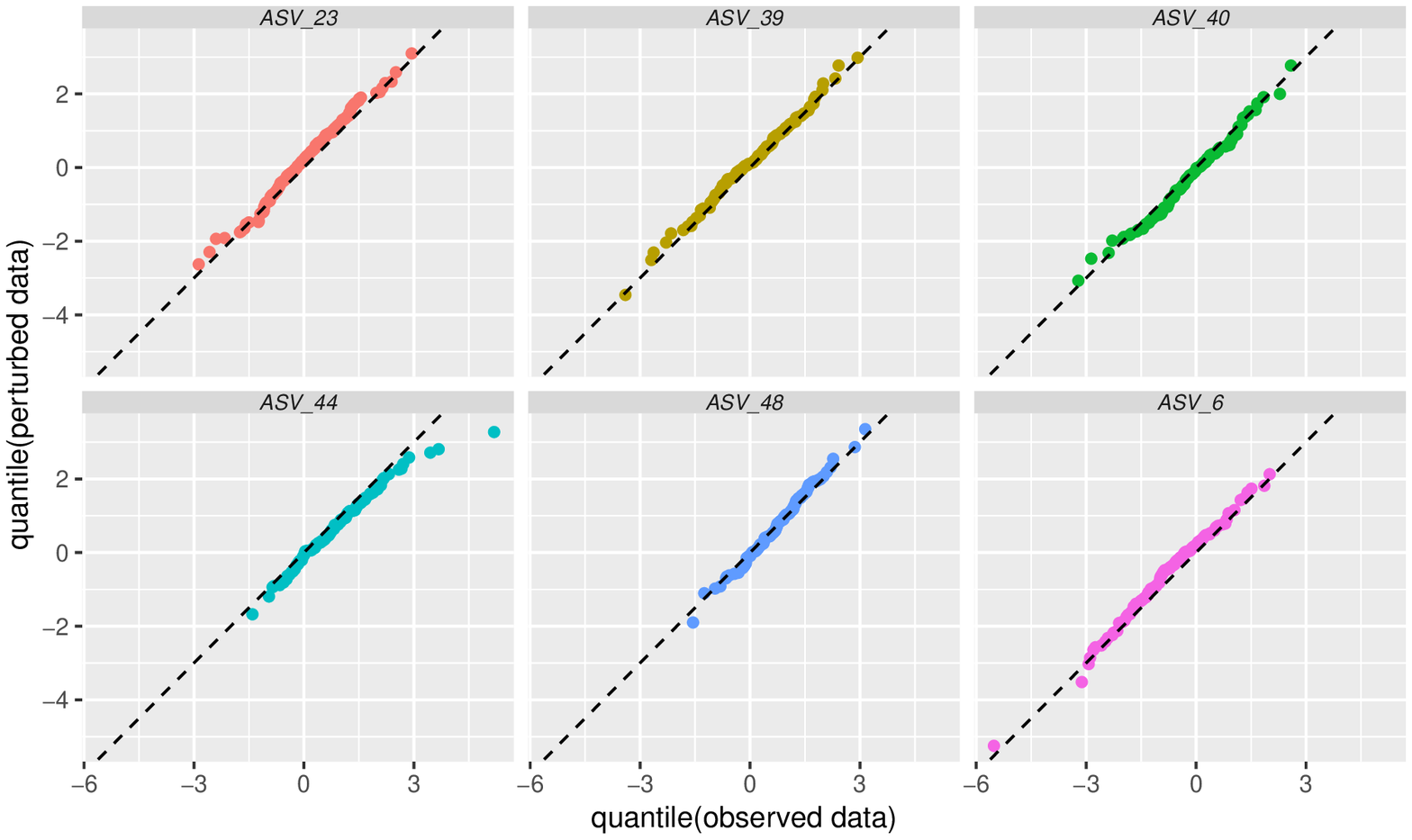}}
		\caption{\label{fig01}Each facet is a sample quantile-quantile plot of test statistic $T_{i}$ for each ASV. Horizontal and vertical axes are the quantiles of $T_{i}$ using observed data and perturbed data, respectively. We simulated the first dataset with realistic parameters and then perturbed the variance of observations to simulate the second set of data. Note that almost all points in each plot fall on the diagonal. This suggests that the sampling distribution of $T_{i}$ agrees in both datasets and thus, $T_{i}$ is a pivotal statistic.}
	\end{figure}
	
	\item Exploratory data analyses (EDA) for choosing initial block size $l_{I}$:
	In practice, we can also choose $l_{I}$ using correlogram. First, we use the variance-stabilizing transformation on the original abundance table. Then, for each ASV, we compute the average transformed abundance at each time point. Finally, we plot within-subject autocorrelation for top six abundances ASVs. These top six ASVs are selected according to the sum of transformed abundances for each ASV. Using the autocorrelation plots, we consider the first occurrence of a lag with sufficiently close to zero autocorrelation as initial block size $l_{I}$. There might be a spurious higher autocorrelation at larger lags due to a small number of observations or abundances close to zero.
\end{itemize}

\subsection*{Computing Resources and extended version of results section:}

We provide \href{https://github.com/PratheepaJ/bootLong_manuscript}{ https://github.com/PratheepaJ/bootLong\_manuscript}, a Github repository to reproduce the results in this manuscript.

\begin{table}
	\caption{\label{tab05} Differentially abundant ASVs at FDR .05 in the Stanford-A cohort data using the \textbf{MBS} method. ASV is identified with species/genus name and strain identification number, lfc is the fold change in $\text{log}_{2}$ scale of abundance in preterm to term subjects, SE is the standard error of lfc, WTS is the Wald test statistic, and p.adj is the adjusted p-value.}
	\fbox{%
		\begin{tabular}{llllll}
		& ASV & lfc & lfcSE & WTS & p.adj \\ 
		\hline
		1 & Gardnerella\_st\_22 & 3.74 & 1.14 & 3.28 & 0.0372 \\ 
		2 & Prevotella\_st\_14 & 2.69 & 0.61 & 4.40 & 0.0012 \\ 
		3 & Lactobacillus gasseri\_st\_53 & -3.28 & 0.99 & -3.32 & 0.0372 \\ 
		\end{tabular}
		}
\end{table}

\begin{figure}
	\centering
	\makebox{\includegraphics[width=\textwidth, scale =1.5]{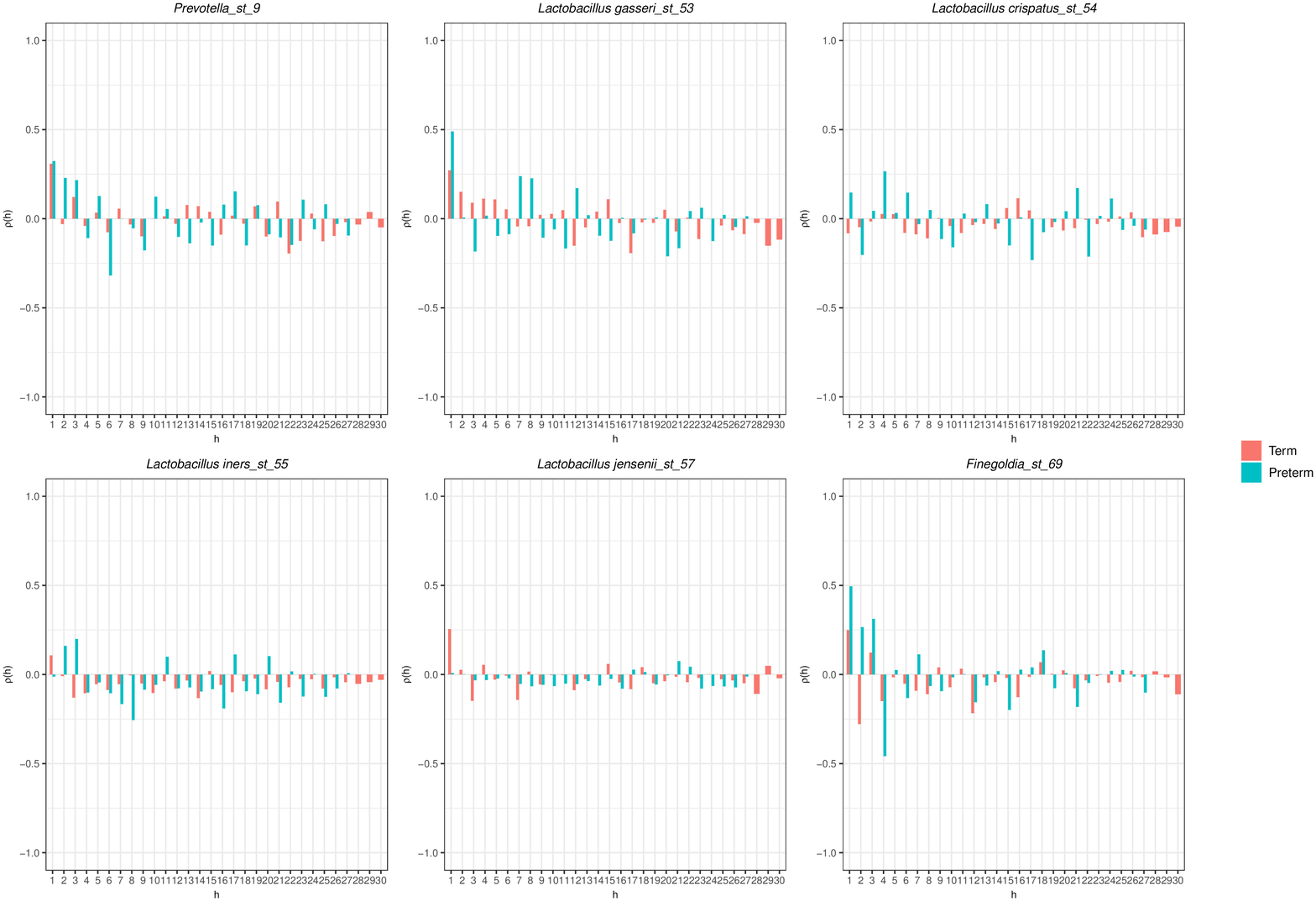}}
	\caption{\label{fig09}Each facet shows a partial autocorrelation (PAC) plot of different ASVs in Stanford-A cohort. Colors refer to the level of a group variable (Preterm/Term). The x-axis label $h$ denotes the lag. The larger spikes are observed at lags less than 8 for both term and preterm. We can choose an initial bock size of 9 because all PAC in all taxa are sufficiently close to zero (PAC $<$ .25) after lag-8. Thus, the initial block size is nine ($l_{I} = 9$).}
\end{figure}

\begin{figure}
	\centering
	\makebox{\includegraphics[width=\textwidth, scale = 1.5]{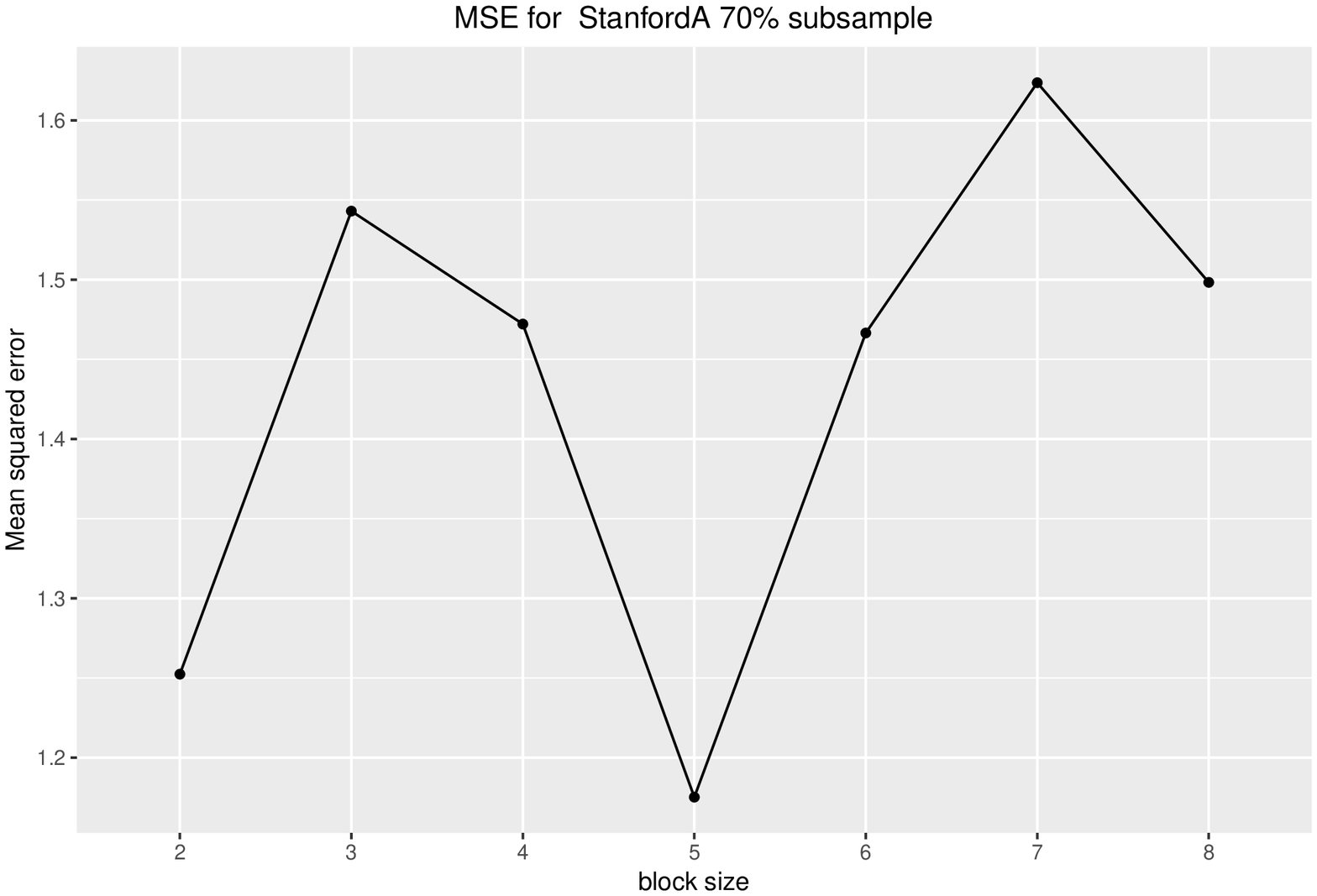}}
	\caption{\label{fig12} The minimum of mean squared error (MSE) occurs at block size five for $70\%$ subsample. Thus, the final optimal block size for the Stanford-A data is five.}
\end{figure}

\begin{table}
	\caption{\label{tab06}Differentially abundant ASVs at FDR .05 in the Stanford-A cohort data using the \textbf{MBB} method. ASV is identified with species/genus name and strain identification number, $\hat{\beta}_{i}$ is the fold change on the arcsinh scale of abundance in preterm to term subjects, p.adj is the adjusted p-value, and CI is the 95\% confidence interval for $\beta_{i}$.}
	\fbox{%
		\begin{tabular}{llllll}
		 & ASV & $\hat{\beta}_{i}$ & lcl & ucl & p.adj \\ 
		 \hline
		 1 & Gardnerella\_st\_22 & 2.58 & 2.10 & 3.08 & $<$.0001 \\ 
		 2 & Atopobium\_st\_89 & 2.31 & 1.73 & 3.36 & $<$.0001 \\ 
		 3 & Prevotella\_st\_11 & 1.63 & 1.21 & 2.27 & 0.0218 \\ 
		 4 & Sneathia\_st\_44 & 1.56 & 1.21 & 1.89 & $<$.0001 \\ 
		 5 & Dialister\_st\_100 & 1.48 & 1.22 & 1.86 & $<$.0001 \\ 
		 6 & Ureaplasma\_st\_50 & 1.31 & 0.96 & 1.73 & $<$.0001 \\ 
		 7 & Prevotella\_st\_17 & 1.30 & 1.01 & 1.65 & $<$.0001 \\ 
		 8 & Finegoldia\_st\_69 & 1.17 & 0.80 & 1.57 & 0.0404 \\ 
		 9 & Clostridium\_st\_86 & 1.08 & 0.79 & 1.47 & $<$.0001 \\ 
		 10 & Corynebacterium\_st\_38 & 0.94 & 0.53 & 1.25 & 0.0218 \\ 
		 11 & Varibaculum\_st\_26 & 0.85 & 0.51 & 1.14 & $<$.0001 \\ 
		 12 & Clostridium\_st\_87 & -0.32 & -0.51 & -0.13 & $<$.0001 \\ 
		 13 & Campylobacter\_st\_2 & -0.35 & -0.56 & -0.10 & 0.0218 \\ 
		 14 & Porphyromonas\_st\_21 & -0.45 & -0.65 & -0.28 & 0.0218 \\ 
		 15 & Sutterella\_st\_5 & -0.48 & -0.68 & -0.30 & $<$.0001 \\ 
		 16 & Propionibacterium acnes\_st\_42 & -0.51 & -0.77 & -0.30 & 0.0218 \\ 
		 17 & F:Propionibacteriaceae\_st\_43 & -0.71 & -0.92 & -0.52 & $<$.0001 \\ 
		 18 & Actinomyces\_st\_28 & -0.79 & -0.99 & -0.56 & $<$.0001 \\ 
		 19 & Pyramidobacter\_st\_47 & -0.85 & -1.21 & -0.64 & $<$.0001 \\ 
		 20 & Porphyromonas\_st\_20 & -0.91 & -1.18 & -0.71 & $<$.0001 \\ 
		 21 & Sutterella\_st\_4 & -0.96 & -1.30 & -0.58 & 0.0404 \\ 
		 22 & F:Lachnospiraceae\_st\_109 & -0.99 & -1.42 & -0.51 & 0.0218 \\ 
		 23 & Lactobacillus iners\_st\_55 & -1.18 & -1.57 & -0.93 & $<$.0001 \\ 
		 24 & F:Dethiosulfovibrionaceae\_st\_46 & -1.21 & -1.61 & -0.84 & 0.0218 \\ 
		 25 & Veillonella\_st\_106 & -1.47 & -2.11 & -0.97 & $<$.0001 \\ 
		 26 & Megasphaera\_st\_104 & -1.50 & -1.74 & -1.28 & $<$.0001 \\ 
		 27 & Clostridium perfringens\_st\_95 & -3.14 & -3.36 & -2.89 & $<$.0001 \\ 
		\end{tabular}}
\end{table}

\begin{figure}
	\centering
	\makebox{\includegraphics[width=\textwidth, scale = 1.5]{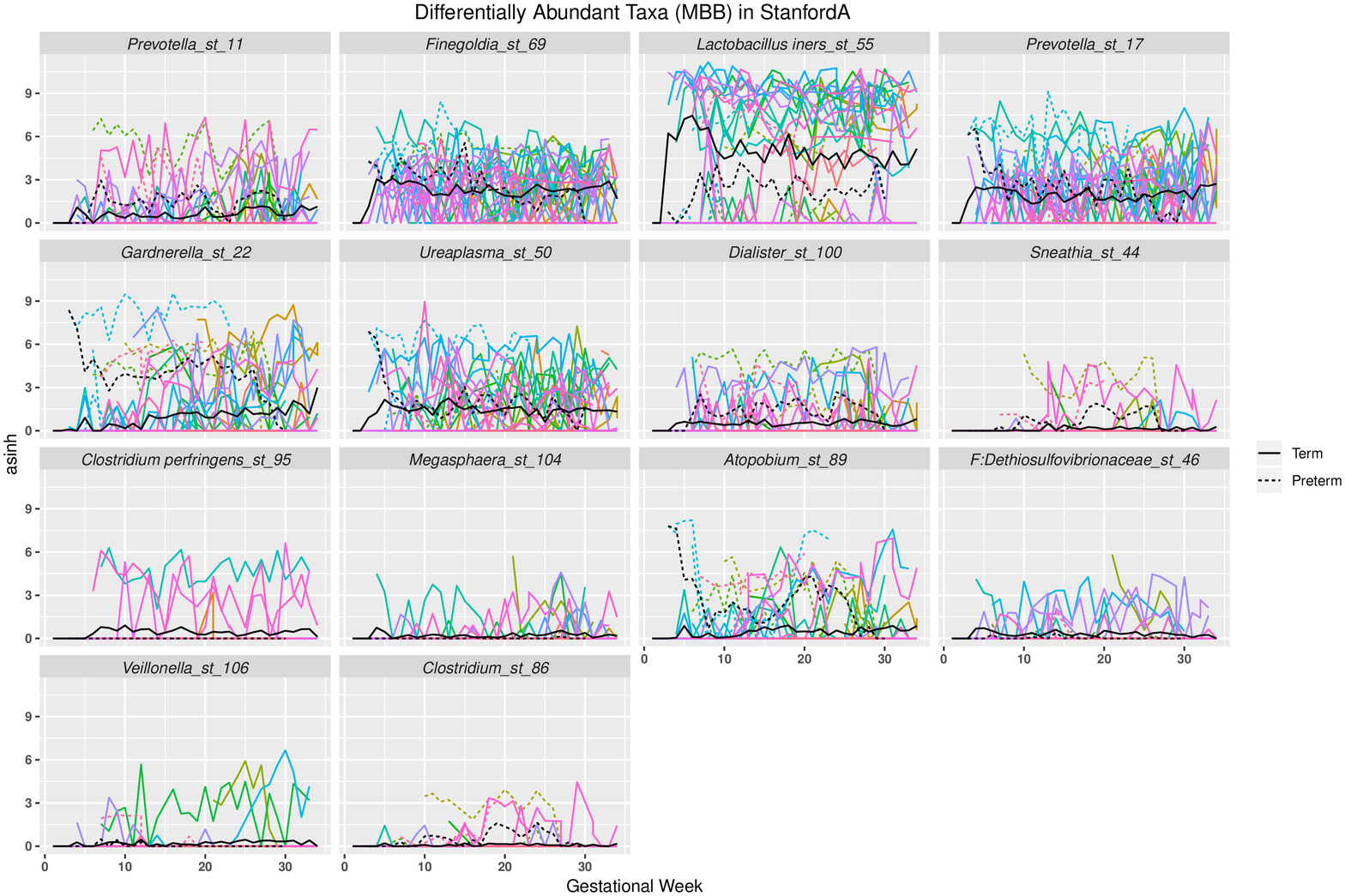}}
	\caption{\label{fig20} arcsinh transformed abundances of significant ASVs with the MBB method and $|\hat{\beta}_{i}| > 1$. The black dotted line and solid line are the mean transformed abundances over time in preterm and term, respectively.}
\end{figure}

\begin{table}
	\caption{\label{tab07}Differentially abundant ASVs at FDR .05 in the Stanford-B cohort data using the \textbf{MBS} method. ASV is identified with species/genus name and strain identification number, lfc is the fold change in $\text{log}_{2}$ scale of abundance in preterm to term subjects, SE is the standard error of lfc, WTS is the Wald test statistic, and p.adj is the adjusted p-value.}
	\centering
	\fbox{%
		\begin{tabular}{llllll}
	& ASV & lfc & lfcSE & WTS & p.adj \\ 
	\hline
	1 & Neisseria\_st\_95 & 3.93 & 0.65 & 6.08 & $<$.0001 \\ 
	2 & Bifidobacterium\_st\_39 & 2.90 & 0.86 & 3.37 & 0.0074 \\ 
	3 & Haemophilus\_st\_65 & 2.60 & 0.62 & 4.23 & 4e-04 \\ 
	4 & Corynebacterium\_1\_st\_81 & 2.43 & 0.79 & 3.07 & 0.015 \\ 
	5 & Fusobacterium\_st\_42 & 2.36 & 0.86 & 2.75 & 0.0304 \\ 
	6 & Blautia\_st\_94 & -1.79 & 0.63 & -2.84 & 0.0254 \\ 
	7 & Porphyromonas\_st\_77 & -1.94 & 0.65 & -2.99 & 0.018 \\ 
	8 & Faecalibacterium\_st\_90 & -2.09 & 0.63 & -3.33 & 0.0077 \\ 
	9 & Pseudobutyrivibrio\_st\_92 & -2.58 & 0.75 & -3.44 & 0.0066 \\ 
	10 & Alloscardovia\_st\_64 & -2.75 & 0.87 & -3.18 & 0.0108 \\ 
	11 & Dialister\_st\_53 & -3.35 & 1.02 & -3.30 & 0.0079 \\ 
	12 & Lactobacillus\_crispatus\_st\_2 & -3.36 & 1.15 & -2.93 & 0.0205 \\ 
	13 & Megasphaera\_st\_4 & -3.65 & 1.06 & -3.43 & 0.0066 \\ 
	14 & Prevotella\_9\_st\_91 & -4.18 & 0.82 & -5.08 & $<$.0001 \\ 
	15 & Atopobium\_st\_10 & -4.37 & 1.24 & -3.52 & 0.0061 \\ 
	16 & Lactobacillus\_jensenii\_st\_5 & -5.37 & 1.20 & -4.47 & 2e-04 \\ 
	17 & Aerococcus\_st\_29 & -6.15 & 1.16 & -5.32 & $<$.0001 \\ 
	18 & Lactobacillus\_crispatus\_st\_8 & -12.83 & 1.54 & -8.33 & $<$.0001 \\ 
	\hline
		\end{tabular}}
\end{table}

\begin{figure}
	\centering
	\begin{tabular}{c}
		\makebox{\includegraphics[width=\textwidth, height = .4\textheight]{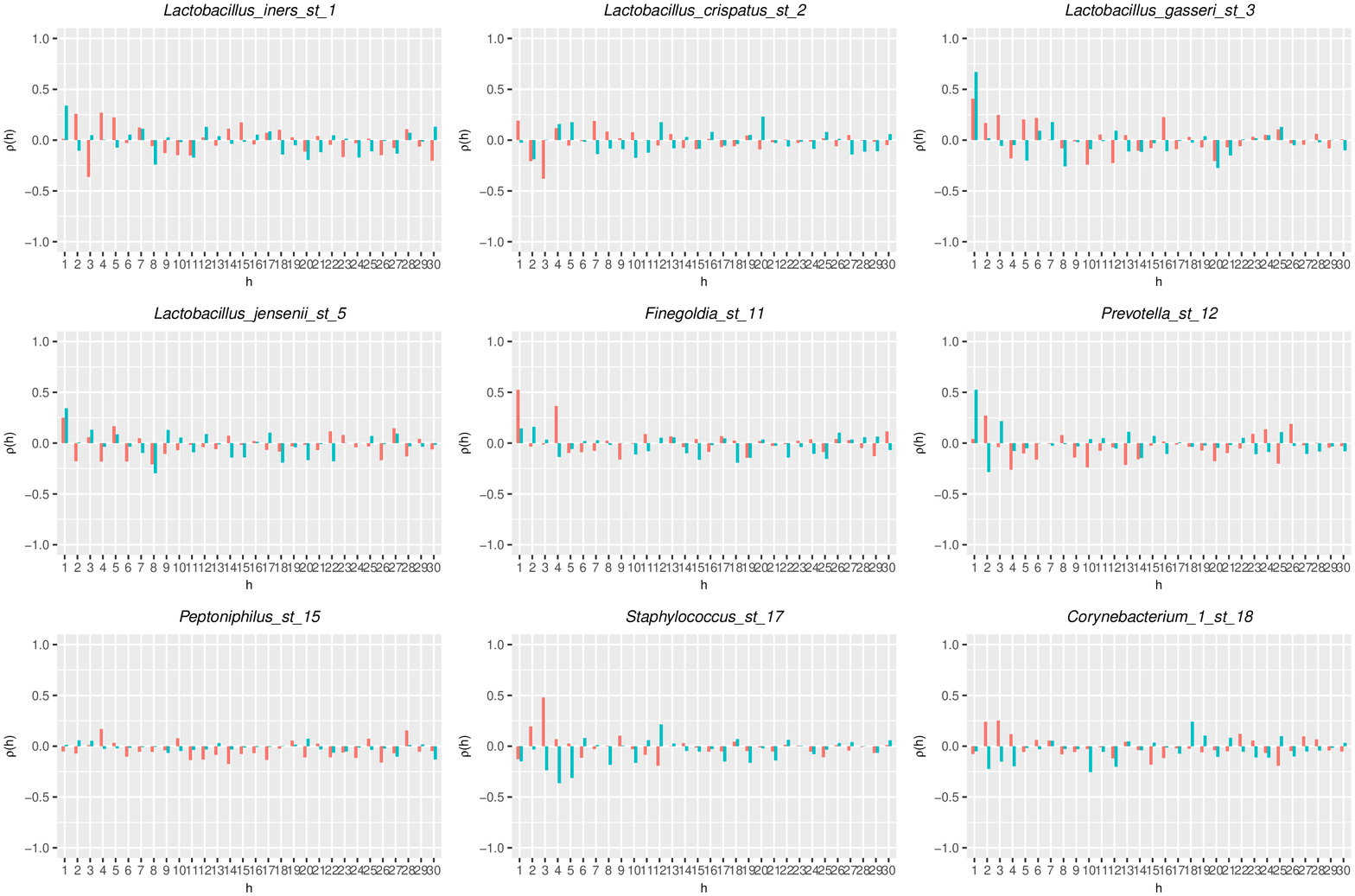}} \\
		\makebox{\includegraphics[width=\textwidth, height = .4\textheight]{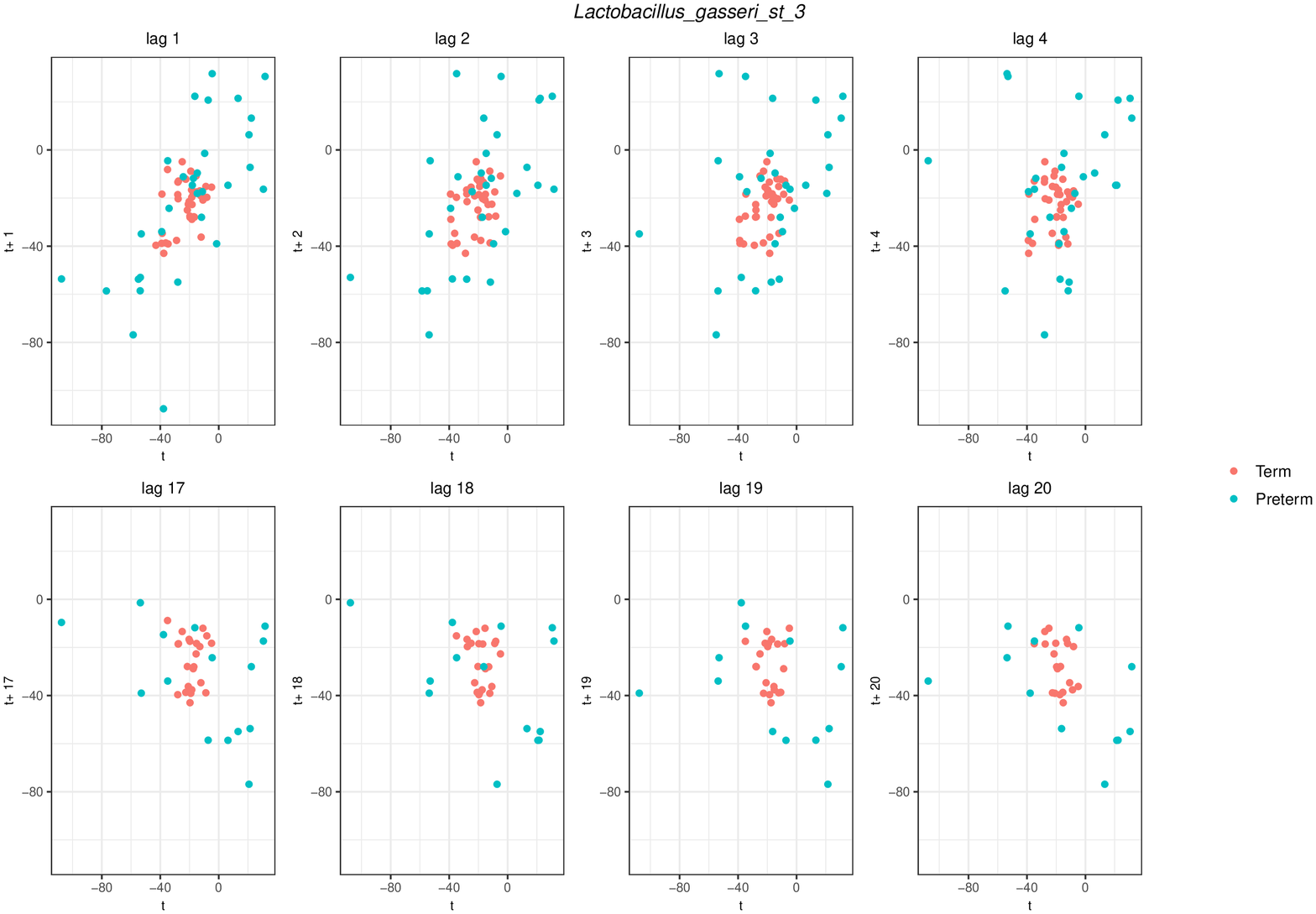} }  
	\end{tabular}
	\caption{ \label{fig10}\textbf{Top:} Each facet shows a partial autocorrelation (PAC) plot of different ASVs in Stanford-B cohort. Colors refer to the level of a group variable (Preterm/Term). The larger spikes (PAC $>$ .25) are observed at lags less than 8 for both term and preterm, except for \text{Lactobacillus\_gasseri\_st\_3} in preterm subjects at lag-20. We can check whether it is a spurious PAC using lag-plots for \textit{Lactobacillus\_gasseri\_st\_3}; \textbf{Bottom:} Lag-20 observations in preterm (green) make less clustering along the diagonal in the negative direction than in lag-1 observations in preterm (green) and there are few observations at this lag. Thus, we considered the larger value of PAC at lag-20 is a spurious effect. We can still choose $l_{I} = 9$.}
\end{figure}

\begin{figure}
	\centering
	\makebox{\includegraphics[width=\textwidth, scale = 1.5]{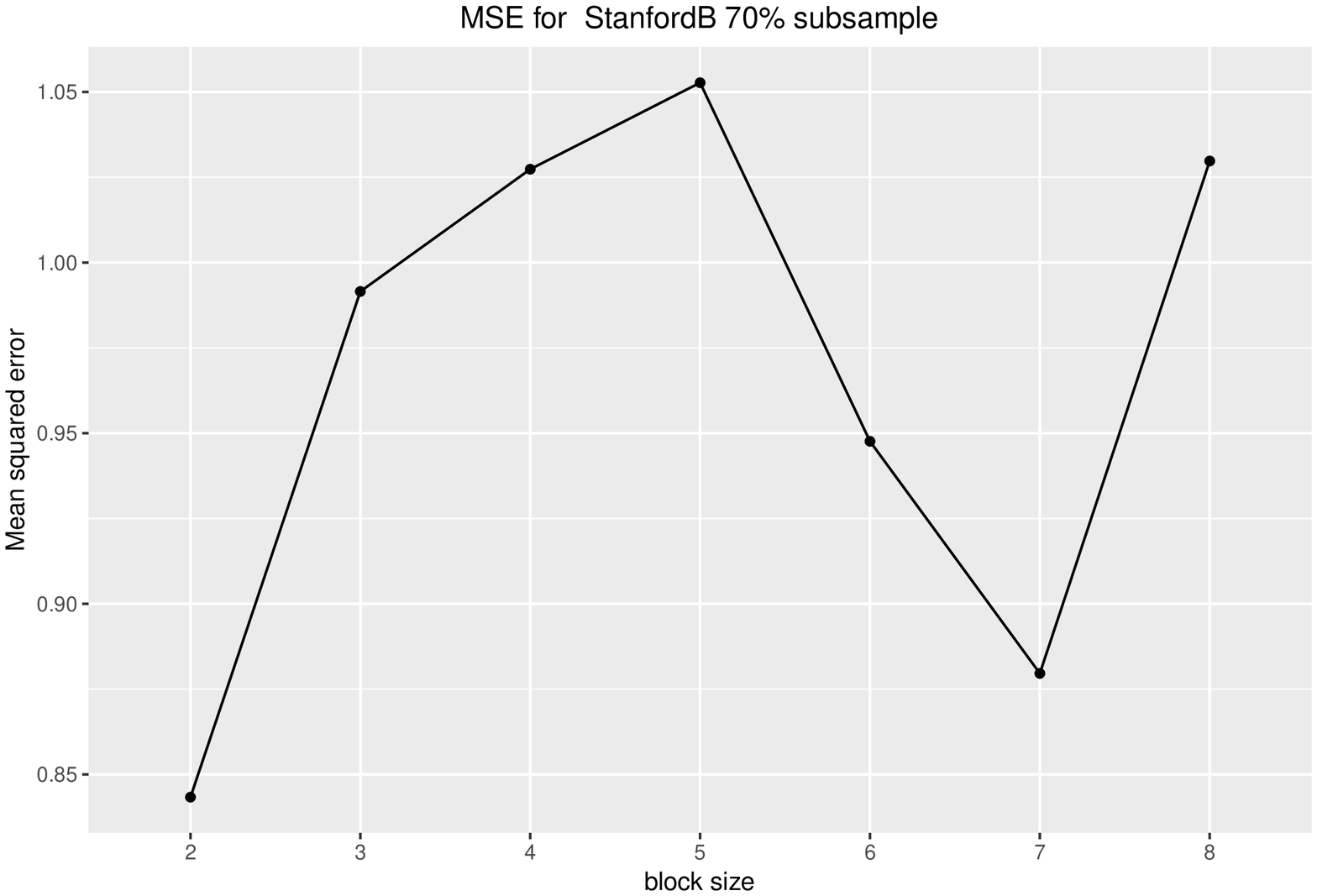}}
	\caption{\label{fig14} The minimum of mean squared error (MSE) occurs at block size two for $70\%$ subsample. According to formula \ref{eq:opt-length-not-eq-length}, the optimal block size for the Stanford-B data is also two.}
\end{figure}

\begin{table}
	\caption{\label{tab08}Differentially abundant ASV at FDR .05 in the Stanford-B cohort data using the \textbf{MBB} method. ASV is identified with species/genus name and strain identification number, $\hat{\beta}_{i}$ is the fold change on the arcsinh scale of abundance in preterm to term subjects, p.adj is the adjusted p-value, and CI is the 95\% confidence interval for $\beta_{i}$.}
	\centering
	\fbox{%
		\begin{tabular}{llllll}
	 & ASV & $\hat{\beta}_{i}$ & lcl & ucl & p.adj \\ 
	 \hline
	1 & Neisseria\_st\_95 & 2.72 & 2.02 & 3.44 & 0.0334 \\ 
	2 & Veillonella\_st\_67 & 2.49 & 1.37 & 3.31 & 0.0187 \\ 
	3 & Aerococcus\_st\_29 & 1.97 & 1.12 & 2.67 & 0.0187 \\ 
	4 & Prevotella\_6\_st\_33 & 1.87 & 1.50 & 2.47 & $<$.0001 \\ 
	5 & Anaerococcus\_st\_61 & 1.74 & 1.33 & 2.11 & $<$.0001 \\ 
	6 & Campylobacter\_st\_56 & 1.63 & 0.99 & 2.49 & 0.0334 \\ 
	7 & Peptoniphilus\_st\_44 & 1.58 & 1.11 & 1.92 & $<$.0001 \\ 
	8 & Dialister\_st\_36 & 1.55 & 1.20 & 1.91 & $<$.0001 \\ 
	9 & Prevotella\_6\_st\_25 & 1.51 & 1.00 & 2.22 & $<$.0001 \\ 
	10 & Gardnerella\_st\_7 & 1.37 & 0.96 & 1.75 & $<$.0001 \\ 
	11 & Peptoniphilus\_st\_63 & 1.37 & 0.94 & 1.81 & 0.0187 \\ 
	12 & Anaerococcus\_st\_51 & 1.35 & 0.61 & 1.78 & 0.0334 \\ 
	13 & Prevotella\_st\_20 & 1.25 & 0.71 & 1.63 & $<$.0001 \\ 
	14 & Murdochiella\_st\_84 & 1.22 & 0.57 & 1.58 & $<$.0001 \\ 
	15 & Prevotella\_st\_41 & 1.18 & 0.71 & 1.61 & $<$.0001 \\ 
	16 & Ezakiella\_st\_26 & 1.17 & 0.31 & 1.62 & 0.0187 \\ 
	17 & Prevotella\_st\_14 & 1.17 & 0.47 & 1.81 & 0.0455 \\ 
	18 & Peptoniphilus\_st\_62 & 1.16 & 0.59 & 1.77 & 0.0455 \\ 
	19 & Howardella\_st\_83 & 1.14 & 0.68 & 1.62 & 0.0455 \\ 
	20 & Lactobacillus\_gasseri\_st\_3 & 0.94 & 0.76 & 1.11 & $<$.0001 \\ 
	21 & Finegoldia\_st\_11 & 0.89 & 0.69 & 1.20 & $<$.0001 \\ 
	22 & Corynebacterium\_1\_st\_19 & 0.80 & 0.31 & 1.26 & 0.0187 \\ 
	23 & Anaerococcus\_st\_22 & 0.79 & 0.54 & 1.05 & $<$.0001 \\ 
	24 & Corynebacterium\_st\_24 & 0.71 & 0.29 & 1.07 & $<$.0001 \\ 
	25 & Actinomyces\_st\_66 & -1.04 & -1.40 & -0.74 & $<$.0001 \\ 
	26 & Lactobacillus\_crispatus\_st\_2 & -1.32 & -1.55 & -1.15 & $<$.0001 \\ 
	27 & Atopobium\_st\_10 & -1.45 & -1.96 & -0.99 & $<$.0001 \\ 
	28 & Lactobacillus\_jensenii\_st\_5 & -1.46 & -1.97 & -1.27 & $<$.0001 \\ 
	29 & Prevotella\_9\_st\_91 & -2.02 & -2.42 & -1.53 & $<$.0001 \\ 
	30 & Lactobacillus\_crispatus\_st\_8 & -9.03 & -9.97 & -8.02 & $<$.0001 \\ 
		\end{tabular}}
\end{table}

\begin{figure}
	\centering
	\makebox{\includegraphics[width=\textwidth, scale =1.5]{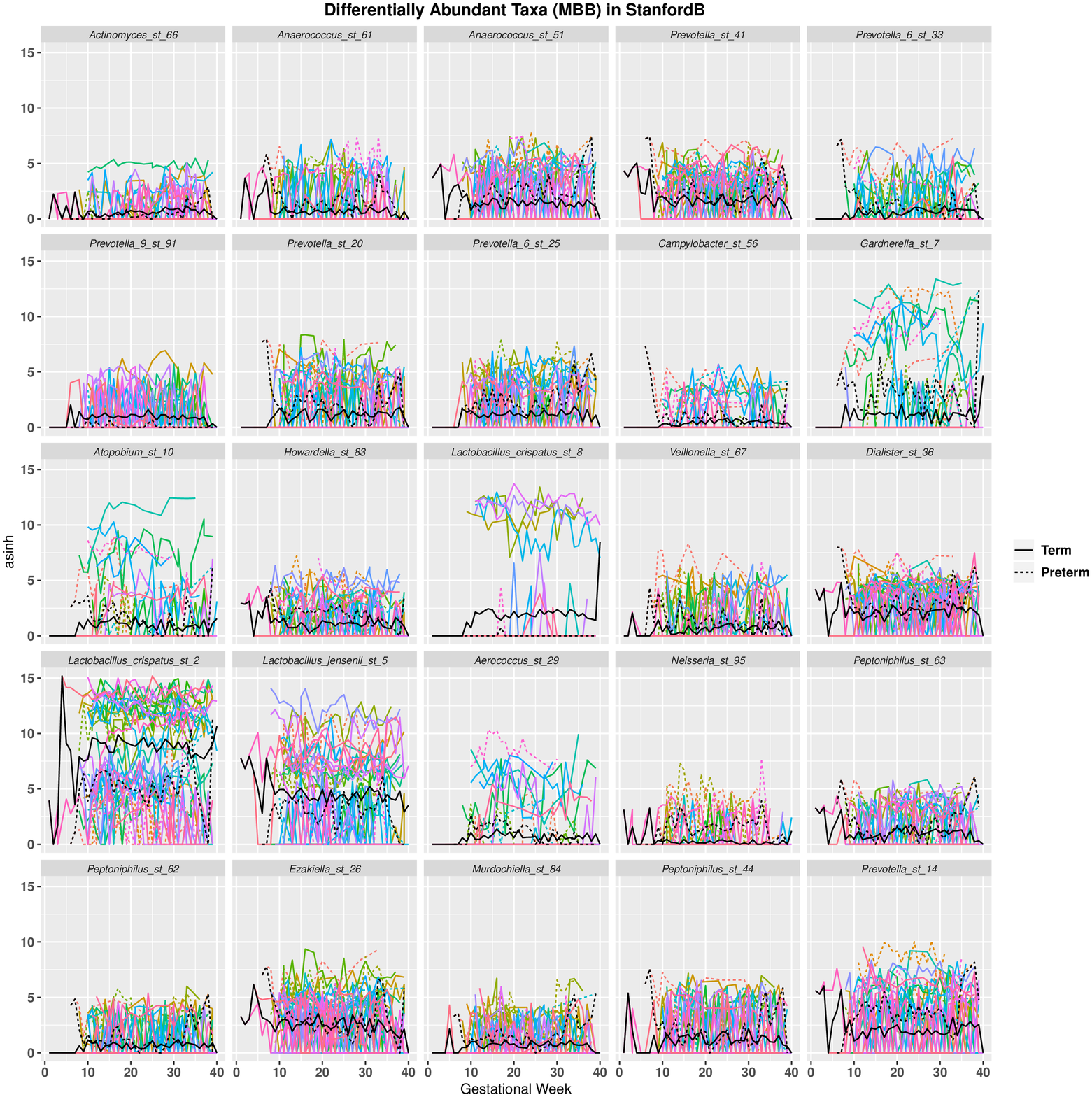}}
	\caption{\label{fig21} arcsinh transformed abundances of significant ASVs in Stanford-B cohort with the MBB method. The black dotted line and solid line are the mean transformed abundances over time in preterm and term, respectively.}
\end{figure}


\begin{table}
	\caption{\label{tab09}Differentially abundant ASVs at FDR .05 in the UAB cohort data using the \textbf{MBS} method. ASV is identified with species/genus name and strain identification number, lfc is the fold change in $\text{log}_{2}$ scale of abundance in preterm to term subjects, SE is the standard error of lfc, WTS is the Wald test statistic, and p.adj is the adjusted p-value.}
	\centering
	\fbox{\begin{tabular}{llllll}
	& ASV & lfc & lfcSE & WTS & p.adj \\ 
	\hline
	1 & Haemophilus\_st\_97 & 3.85 & 0.55 & 7.00 & $<$.0001 \\ 
	2 & Prevotella\_st\_36 & 3.09 & 0.69 & 4.50 & 1e-04 \\ 
	3 & Prevotella\_st\_10 & 3.03 & 0.69 & 4.40 & 1e-04 \\ 
	4 & Streptococcus\_st\_69 & 2.87 & 0.49 & 5.87 & $<$.0001 \\ 
	5 & Veillonella\_st\_99 & 2.58 & 0.44 & 5.81 & $<$.0001 \\ 
	6 & Sneathia\_st\_18 & 2.49 & 0.62 & 4.03 & 5e-04 \\ 
	7 & Fastidiosipila\_st\_180 & 2.44 & 0.45 & 5.42 & $<$.0001 \\ 
	8 & Prevotella\_st\_39 & 2.32 & 0.69 & 3.37 & 0.0041 \\ 
	9 & Bifidobacterium\_st\_54 & 2.26 & 0.49 & 4.62 & 1e-04 \\ 
	10 & Sutterella\_st\_105 & 2.21 & 0.42 & 5.26 & $<$.0001 \\ 
	11 & Gemella\_st\_64 & 2.18 & 0.52 & 4.20 & 3e-04 \\ 
	12 & Gardnerella\_st\_35 & 2.15 & 0.60 & 3.58 & 0.0022 \\ 
	13 & Sneathia\_st\_13 & 2.15 & 0.62 & 3.47 & 0.0029 \\ 
	14 & Prevotella\_st\_53 & 2.05 & 0.43 & 4.72 & $<$.0001 \\ 
	15 & Campylobacter\_st\_38 & 1.96 & 0.44 & 4.48 & 1e-04 \\ 
	16 & Mycoplasma\_st\_19 & 1.91 & 0.62 & 3.06 & 0.0102 \\ 
	17 & Arcanobacterium\_st\_168 & 1.87 & 0.39 & 4.74 & $<$.0001 \\ 
	18 & Dialister\_st\_135 & 1.84 & 0.50 & 3.66 & 0.0019 \\ 
	19 & Prevotella\_6\_st\_32 & 1.81 & 0.50 & 3.62 & 0.0022 \\ 
	20 & Senegalimassilia\_st\_48 & 1.75 & 0.53 & 3.32 & 0.0047 \\ 
	21 & Porphyromonas\_st\_115 & 1.64 & 0.44 & 3.70 & 0.0017 \\ 
	22 & Dialister\_st\_79 & 1.63 & 0.51 & 3.22 & 0.0063 \\ 
	23 & Dialister\_st\_50 & 1.53 & 0.35 & 4.40 & 1e-04 \\ 
	24 & Prevotella\_7\_st\_142 & 1.46 & 0.39 & 3.78 & 0.0014 \\ 
	25 & Escherichia/Shigella\_st\_45 & 1.40 & 0.53 & 2.63 & 0.0306 \\ 
	26 & Lactobacillus\_coleohominis\_st\_89 & 1.39 & 0.56 & 2.49 & 0.0406 \\ 
	27 & Gemella\_st\_114 & 1.37 & 0.46 & 2.97 & 0.0129 \\ 
	28 & Fusobacterium\_st\_59 & 1.32 & 0.52 & 2.56 & 0.0357 \\ 
	29 & Dialister\_st\_30 & 1.29 & 0.41 & 3.11 & 0.0087 \\ 
	30 & Moryella\_st\_173 & 1.27 & 0.34 & 3.72 & 0.0016 \\ 
\vdots \\
	45 & Prevotella\_7\_st\_117 & -1.32 & 0.51 & -2.60 & 0.0324 \\ 
	46 & Corynebacterium\_st\_126 & -1.40 & 0.40 & -3.47 & 0.0029 \\ 
	47 & Staphylococcus\_st\_106 & -1.46 & 0.52 & -2.79 & 0.0208 \\ 
	48 & Anaerococcus\_st\_87 & -1.75 & 0.49 & -3.60 & 0.0022 \\ 
	49 & Brevibacterium\_st\_81 & -1.85 & 0.40 & -4.67 & 1e-04 \\ 
	50 & Lactobacillus\_gasseri\_st\_4 & -1.94 & 0.61 & -3.17 & 0.0074 \\ 
	51 & Enterococcus\_st\_104 & -2.21 & 0.48 & -4.60 & 1e-04 \\ 
	52 & Sutterella\_st\_131 & -2.27 & 0.47 & -4.81 & $<$.0001 \\ 
	53 & Corynebacterium\_1\_st\_102 & -2.55 & 0.44 & -5.79 & $<$.0001 \\ 
		\end{tabular}}
\end{table}

\begin{figure}
	\centering
	\begin{tabular}{c}
		\makebox{\includegraphics[width=\textwidth, , height = .4\textheight]{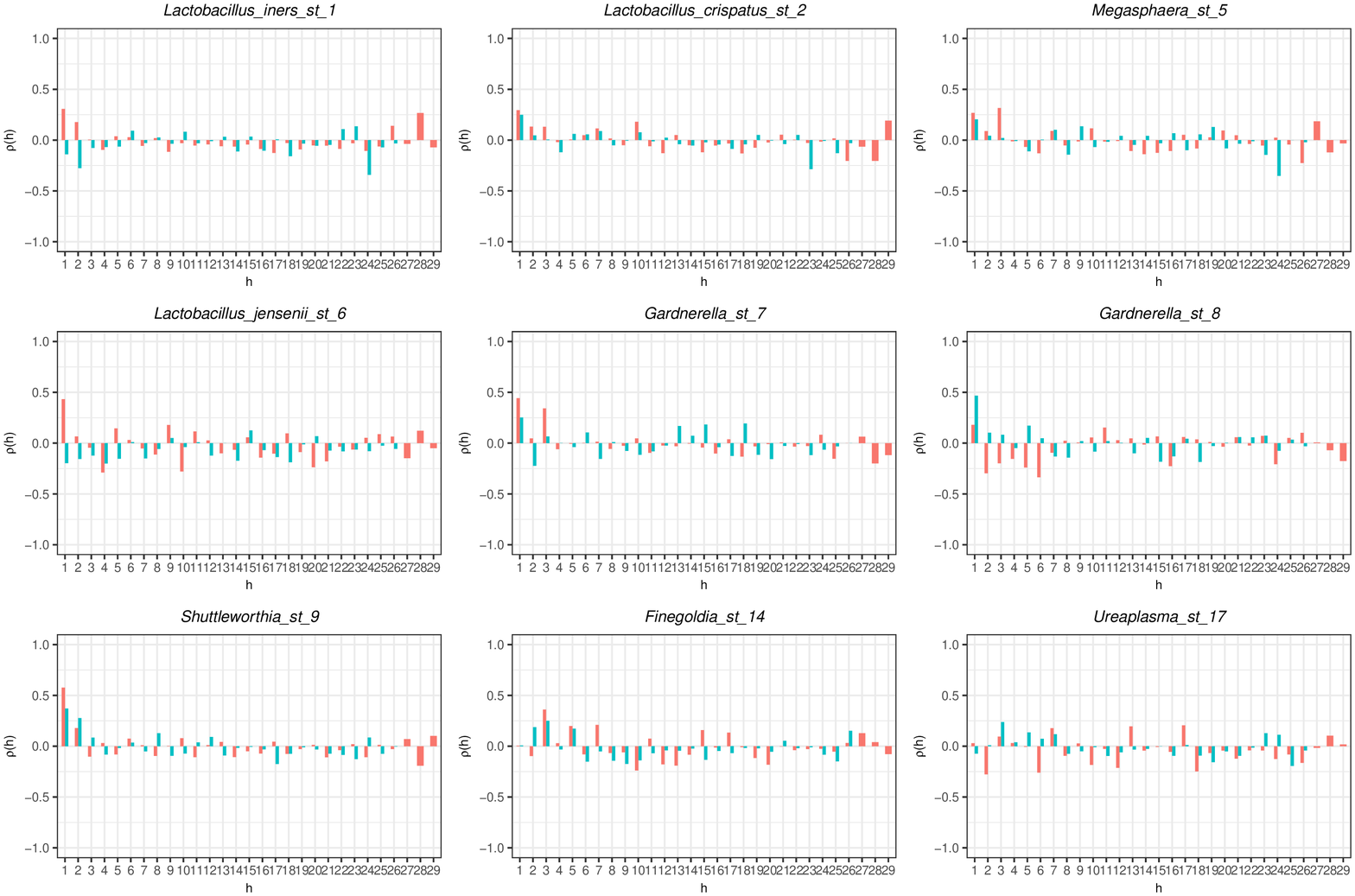} }\\
		\makebox{\includegraphics[width=\textwidth, , height = .4\textheight]{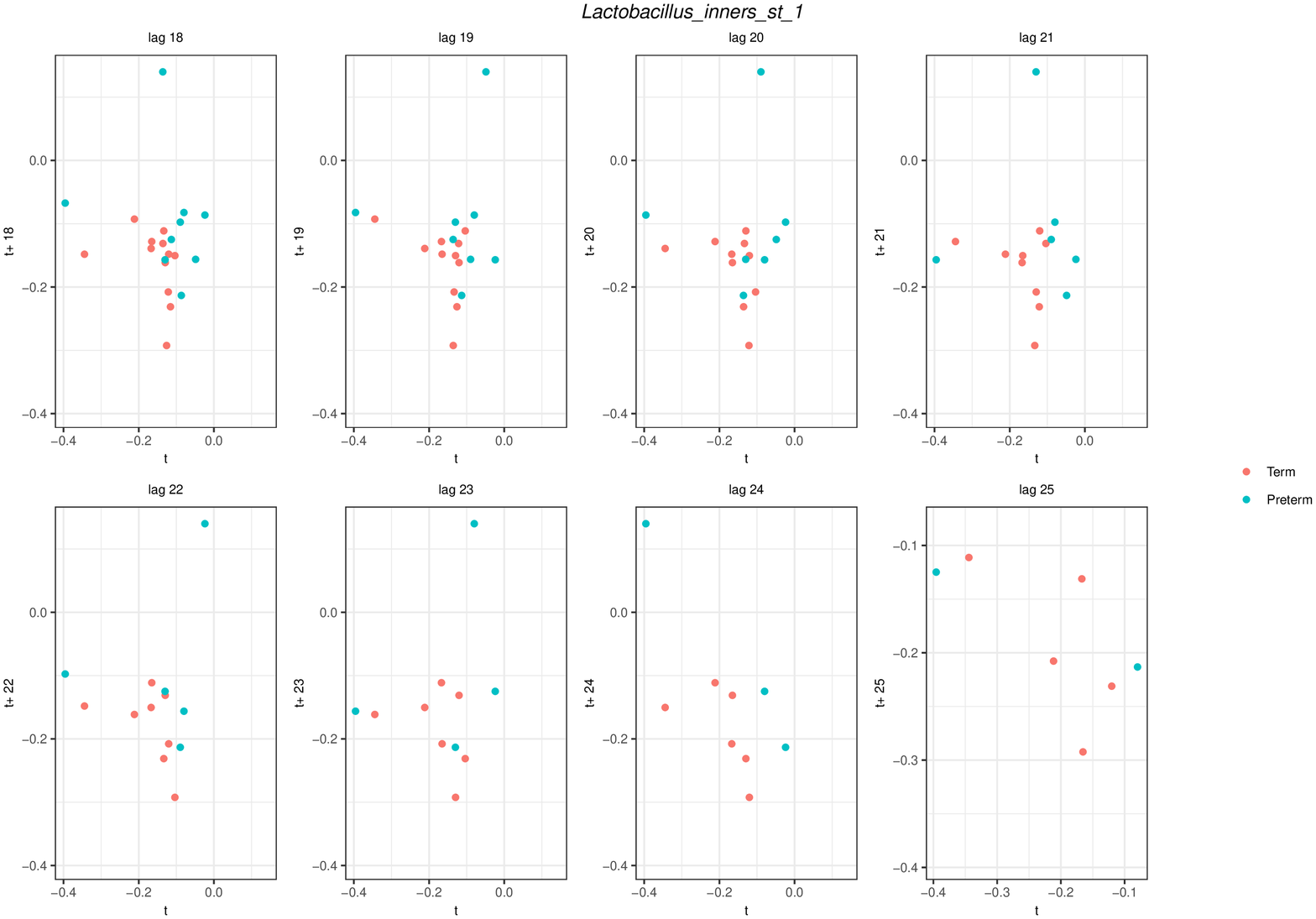}}   
	\end{tabular}
	\caption{ \label{fig11}\textbf{Top:} Each facet shows a partial autocorrelation (PAC) plot of different ASVs in UAB cohort. Colors refer to the level of a group variable (Preterm/Term). The x-axis label $h$ denotes the lag. The larger spikes (PAC $>$ .25) are observed at lags less than 10 for both term and preterm, except for \textit{Lactobacillus inners\_st\_1} in preterm subjects at lag-24. We can check whether it is a spurious PAC using lag-plot; \textbf{Bottom:} There are only three lag-24 observations of \textit{Lactobacillus inners\_st\_1} in preterm (green). Thus, we consider the larger value of PAC at lag-24 is a spurious effect. We can still choose $l_{I} = 11$.}
\end{figure}

\begin{figure}
	\centering
	\makebox{\includegraphics[width=\textwidth, scale = 1.5]{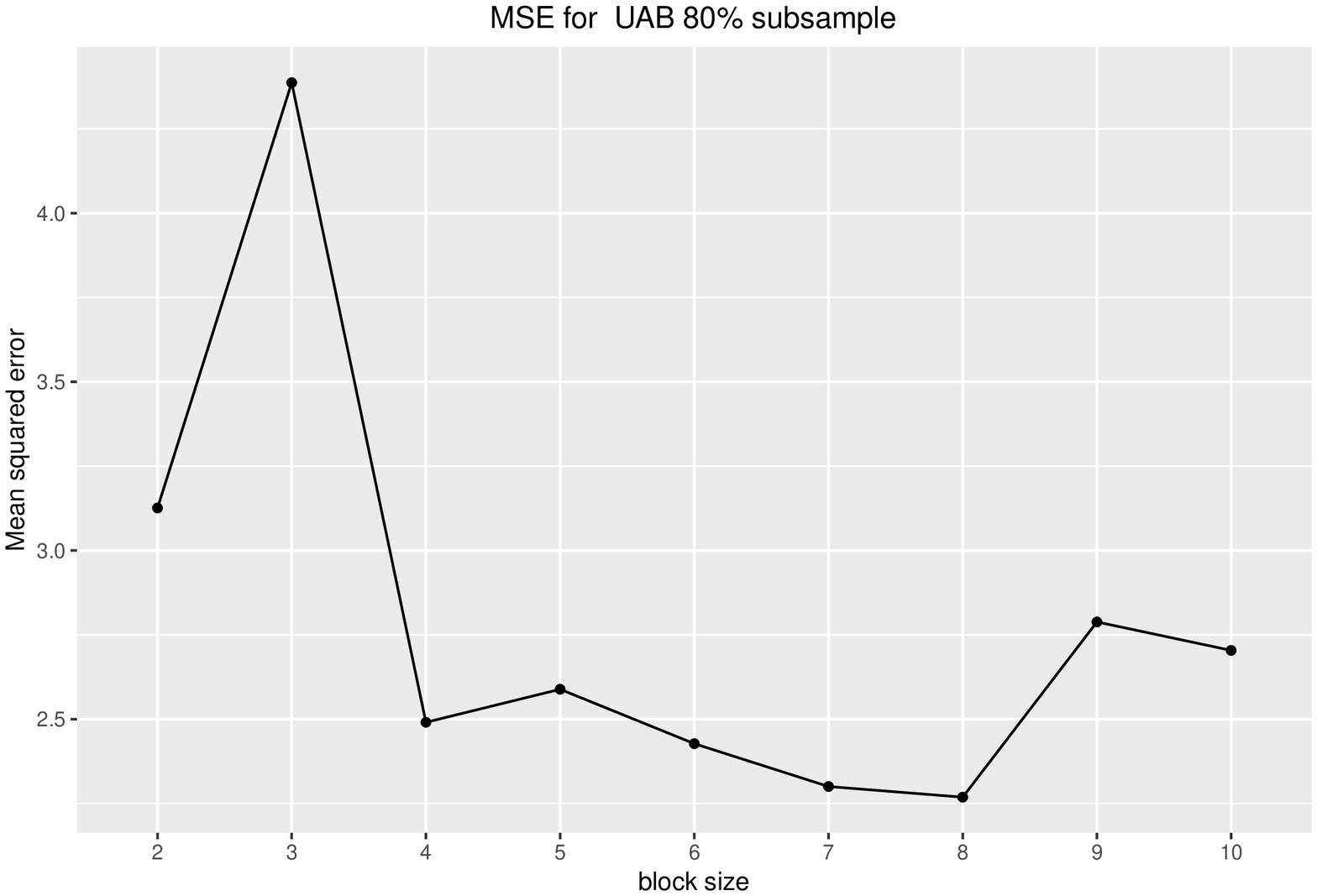}}
	\caption{\label{fig13} The minimum of mean squared error (MSE) occurs at block size eight for $80\%$ subsample. The optimal block size for the UAB data is eight.}
\end{figure}

\begin{table}
	\caption{\label{tab10}Differentially abundant ASVs at FDR .05 in the UAB cohort data using the MBB method. ASV is identified with species/genus name and strain identification number, $\hat{\beta}_{i}$ is the fold change on the arcsinh scale of abundance in preterm to term subjects, p.adj is the adjusted p-value, and CI is the 95\% confidence interval for $\beta_{i}$.}
	\centering
	\fbox{\begin{tabular}{llllll}
		 & ASV &$\hat{\beta}_{i}$ & lcl & ucl & p.adj \\ 
		 \hline
		1 & Streptococcus\_st\_69 & 1.65 & 1.13 & 2.46 & $<$.0001 \\ 
		2 & Veillonella\_st\_99 & 1.63 & 0.93 & 2.18 & $<$.0001 \\ 
		3 & Arcanobacterium\_st\_168 & 1.35 & 0.90 & 1.89 & 0.0218 \\ 
		4 & Prevotella\_7\_st\_55 & 1.07 & 0.39 & 1.47 & $<$.0001 \\ 
		5 & Sneathia\_st\_18 & 1.07 & 0.95 & 1.26 & $<$.0001 \\ 
		6 & Prevotella\_st\_36 & 0.63 & 0.26 & 1.02 & 0.0218 \\ 
		7 & Corynebacterium\_1\_st\_37 & 0.60 & 0.29 & 0.75 & 0.0218 \\ 
		8 & Prevotella\_st\_39 & 0.58 & 0.38 & 0.76 & $<$.0001 \\ 
		9 & Sneathia\_st\_13 & 0.58 & 0.46 & 0.71 & $<$.0001 \\ 
		10 & Prevotella\_6\_st\_32 & 0.56 & 0.22 & 0.85 & $<$.0001 \\ 
		11 & Fastidiosipila\_st\_180 & 0.56 & 0.38 & 0.70 & $<$.0001 \\ 
\vdots \\
		17 & Ezakiella\_st\_33 & -0.46 & -0.68 & -0.26 & 0.0218 \\ 
		18 & Ruminococcaceae\_UCG-014\_st\_124 & -0.50 & -0.79 & -0.17 & 0.0218 \\ 
		19 & Peptoniphilus\_st\_163 & -0.50 & -0.71 & -0.36 & 0.0218 \\ 
		20 & Actinomyces\_st\_177 & -0.53 & -0.74 & -0.38 & 0.0381 \\ 
		21 & Helcococcus\_st\_139 & -0.58 & -0.92 & -0.32 & 0.0381 \\ 
		22 & Peptoniphilus\_st\_94 & -0.61 & -0.86 & -0.31 & 0.0218 \\ 
		23 & Corynebacterium\_st\_31 & -0.61 & -0.90 & -0.39 & 0.0381 \\ 
		24 & Anaerococcus\_st\_41 & -0.73 & -1.03 & -0.47 & 0.0218 \\ 
		25 & Peptoniphilus\_st\_66 & -0.80 & -1.16 & -0.57 & $<$.0001 \\ 
		26 & Anaerococcus\_st\_75 & -0.81 & -1.01 & -0.55 & $<$.0001 \\ 
		27 & Atopobium\_st\_72 & -0.83 & -1.04 & -0.65 & $<$.0001 \\ 
		28 & Gardnerella\_st\_8 & -0.83 & -0.97 & -0.70 & $<$.0001 \\ 
		29 & Prevotella\_st\_68 & -0.84 & -1.23 & -0.60 & 0.0218 \\ 
		30 & Peptoniphilus\_st\_158 & -0.85 & -1.21 & -0.72 & $<$.0001 \\ 
		31 & Negativicoccus\_st\_175 & -0.87 & -1.31 & -0.57 & 0.0218 \\ 
		32 & Porphyromonas\_st\_109 & -0.87 & -1.37 & -0.66 & 0.0381 \\ 
		33 & Prevotella\_st\_132 & -0.88 & -1.08 & -0.57 & 0.0218 \\ 
		34 & Prevotella\_st\_57 & -0.94 & -1.31 & -0.77 & $<$.0001 \\ 
		35 & Lactobacillus\_gasseri\_st\_4 & -0.94 & -1.17 & -0.70 & $<$.0001 \\ 
		36 & Anaerococcus\_st\_110 & -0.97 & -1.31 & -0.81 & $<$.0001 \\ 
		37 & Dermabacter\_st\_150 & -1.00 & -1.37 & -0.71 & 0.0218 \\ 
		38 & Anaerococcus\_st\_87 & -1.11 & -1.44 & -0.71 & 0.0218 \\ 
		39 & Corynebacterium\_1\_st\_52 & -1.13 & -1.41 & -0.97 & $<$.0001 \\ 
		40 & Atopobium\_st\_12 & -1.28 & -1.48 & -0.98 & $<$.0001 \\ 
		41 & Corynebacterium\_1\_st\_40 & -1.31 & -1.65 & -1.08 & $<$.0001 \\ 
		42 & Gallicola\_st\_82 & -1.42 & -1.80 & -1.17 & $<$.0001 \\ 
		43 & Methylobacterium\_st\_162 & -1.47 & -2.16 & -0.94 & 0.0381 \\ 
		44 & Ureaplasma\_st\_17 & -1.55 & -1.89 & -1.23 & 0.0218 \\ 
		45 & Corynebacterium\_1\_st\_76 & -1.87 & -2.38 & -1.53 & $<$.0001 \\ 
		\hline
		\end{tabular}}
\end{table}

\begin{figure}
	\centering
	\makebox{\includegraphics[width=\textwidth, scale = 1.5]{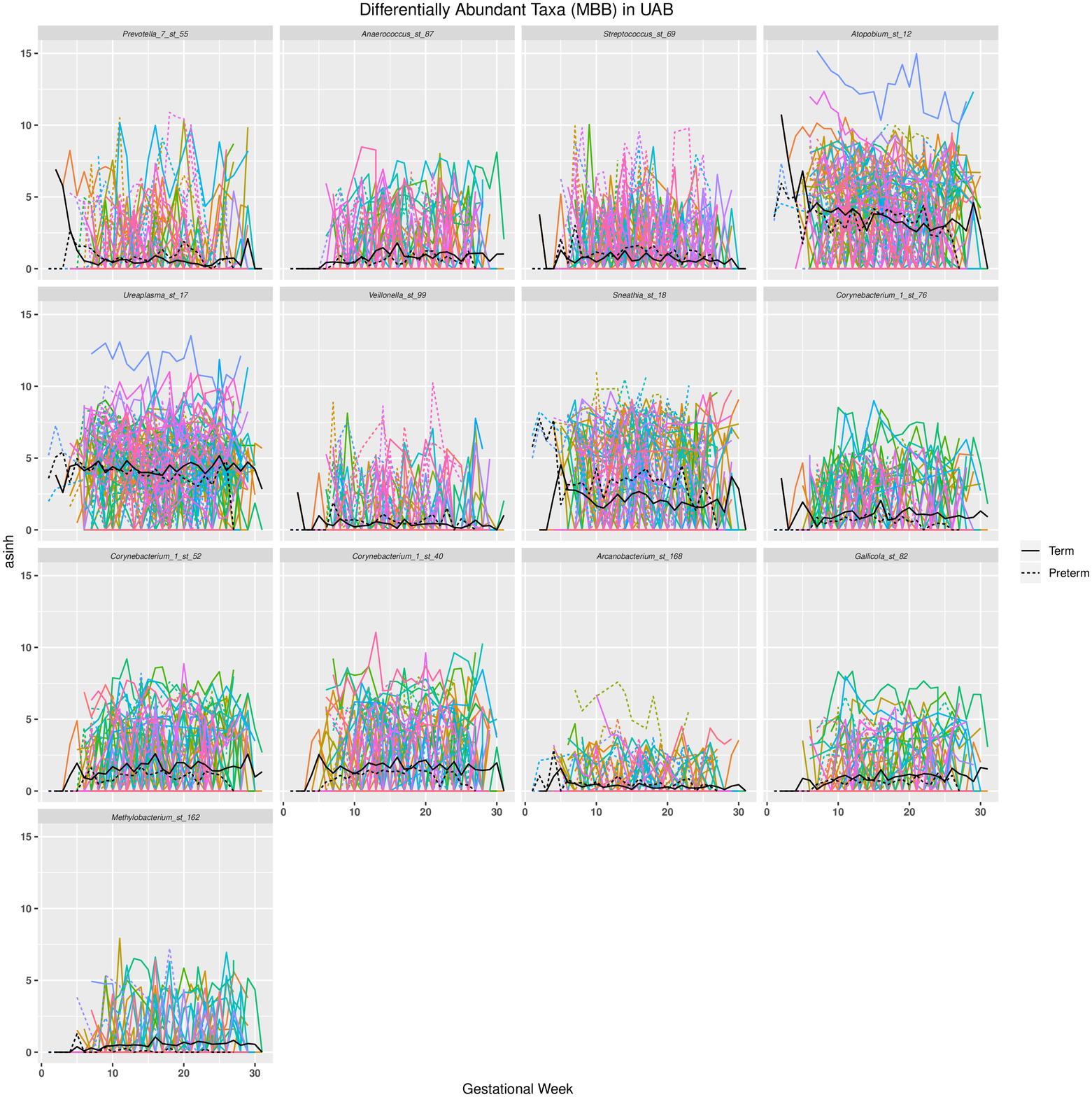}}
	\caption{\label{fig22} arcsinh transformed abundances of significant ASVs in UAB cohort with the MBB method and $|\hat{\beta}_{i}| > 1$. The black dotted line and solid line are the mean transformed abundances over time in preterm and term, respectively.}
\end{figure}


\begin{figure}
	\centering
	\makebox{\includegraphics[width=\textwidth, scale = 1.5]{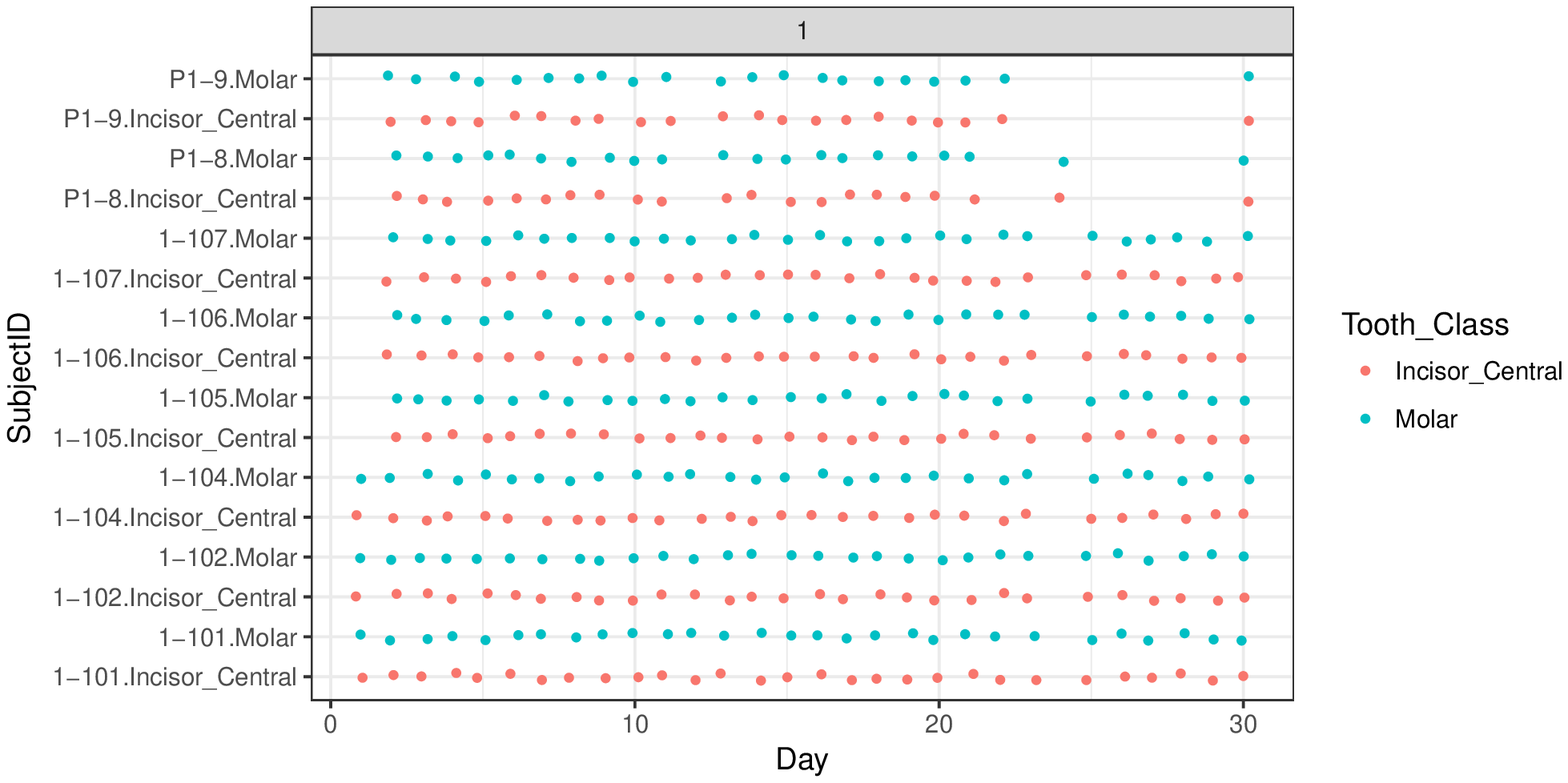}}
	\caption{\label{fig15}Sampling schedule for oral supragingival plaque samples (upper jaw) of lingual surface from teeth 3 \& 14 molars and 8 \& 9 incisors.}
\end{figure}

\begin{table}
	\caption{\label{tab11}Differentially abundant ASVs at FDR .05 in the oral data using the \textbf{MBS} method. ASV is identified with species/genus name and strain identification number, lfc is the fold change in $\text{log}_{2}$ scale of abundance in molar to incisors tooth classes, SE is the standard error of lfc, WTS is the Wald test statistic, and p.adj is the adjusted p-value.}
	\centering
	\fbox{\begin{tabular}{llllll}
			& ASV & lfc & lfcSE & WTS & p.adj \\ 
			\hline
			1 & Prevotella\_denticola\_st\_74 & 3.45 & 0.93 & 3.73 & 0.0012 \\ 
			2 & Prevotella\_histicola\_st\_36 & 3.06 & 1.10 & 2.78 & 0.0156 \\ 
			3 & Veillonella\_st\_28 & 3.02 & 0.98 & 3.06 & 0.0088 \\ 
			4 & Prevotella\_salivae\_st\_47 & 2.98 & 0.78 & 3.80 & 0.001 \\ 
			5 & Campylobacter\_concisus\_st\_55 & 2.71 & 0.75 & 3.61 & 0.0018 \\ 
			6 & Prevotella\_nigrescens\_st\_49 & 2.60 & 0.76 & 3.44 & 0.0028 \\ 
			7 & Prevotella\_pallens\_st\_70 & 2.52 & 0.86 & 2.92 & 0.0131 \\ 
			8 & Oribacterium\_sinus\_st\_71 & 2.39 & 0.67 & 3.58 & 0.002 \\ 
			9 & Prevotella\_melaninogenica\_st\_37 & 2.10 & 0.62 & 3.36 & 0.0035 \\ 
			10 & Dialister\_invisus\_st\_78 & 1.98 & 0.73 & 2.71 & 0.0192 \\ 
			11 & Prevotella\_nanceiensis\_st\_51 & 1.77 & 0.62 & 2.88 & 0.0133 \\ 
			12 & Prevotella\_st\_76 & 1.65 & 0.69 & 2.41 & 0.0412 \\ 
			13 & Catonella\_morbi\_st\_82 & -1.60 & 0.68 & -2.34 & 0.0485 \\ 
			14 & Brachymonas\_st\_45 & -2.21 & 0.88 & -2.52 & 0.0309 \\ 
			15 & Capnocytophaga\_gingivalis\_st\_32 & -2.49 & 0.87 & -2.85 & 0.0133 \\ 
			16 & Leptotrichia\_st\_33 & -2.52 & 0.88 & -2.87 & 0.0133 \\ 
			17 & Abiotrophia\_defectiva\_st\_9 & -2.55 & 0.73 & -3.50 & 0.0025 \\ 
			18 & Kingella\_oralis\_st\_31 & -2.76 & 0.96 & -2.87 & 0.0133 \\ 
			19 & Capnocytophaga\_leadbetteri\_st\_43 & -2.98 & 0.86 & -3.47 & 0.0026 \\ 
			20 & Neisseria\_st\_4 & -3.01 & 0.93 & -3.24 & 0.0052 \\ 
			21 & Actinomyces\_st\_39 & -3.24 & 1.12 & -2.89 & 0.0133 \\ 
			22 & Actinomyces\_massiliensis\_st\_34 & -3.35 & 1.10 & -3.05 & 0.0088 \\ 
			23 & Actinomyces\_st\_10 & -3.36 & 0.67 & -5.00 & $<$.0001 \\ 
			24 & Capnocytophaga\_sputigena\_st\_48 & -3.61 & 0.78 & -4.62 & $<$.0001 \\ 
			25 & Streptococcus\_sanguinis\_st\_5 & -3.81 & 0.57 & -6.70 & $<$.0001 \\ 
			26 & Rothia\_dentocariosa\_st\_1 & -4.20 & 0.60 & -7.02 & $<$.0001 \\ 
			27 & Cardiobacterium\_hominis\_st\_24 & -4.28 & 0.89 & -4.83 & $<$.0001 \\ 
			28 & Rothia\_st\_11 & -4.67 & 0.69 & -6.73 & $<$.0001 \\ 
			29 & Neisseria\_st\_26 & -5.14 & 0.84 & -6.09 & $<$.0001 \\ 
			30 & Corynebacterium\_durum\_st\_16 & -5.68 & 0.80 & -7.13 & $<$.0001 \\ 
		\end{tabular}}
	\end{table}

\begin{figure}
	\centering
	\begin{tabular}{c}
		\makebox{\includegraphics[width=\textwidth,  height = .4\textheight]{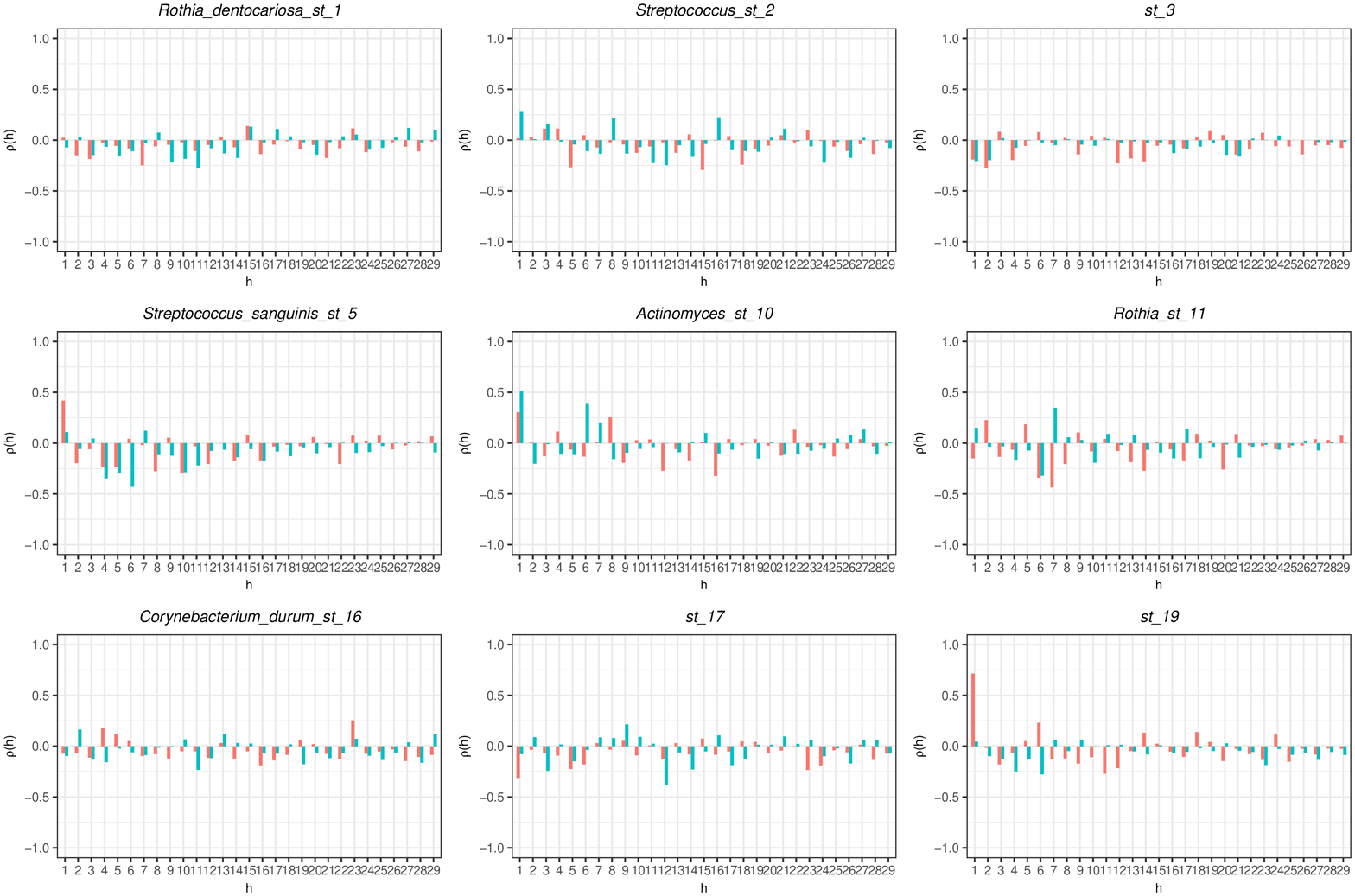}} \\
		\makebox{\includegraphics[width=\textwidth, height = .4\textheight]{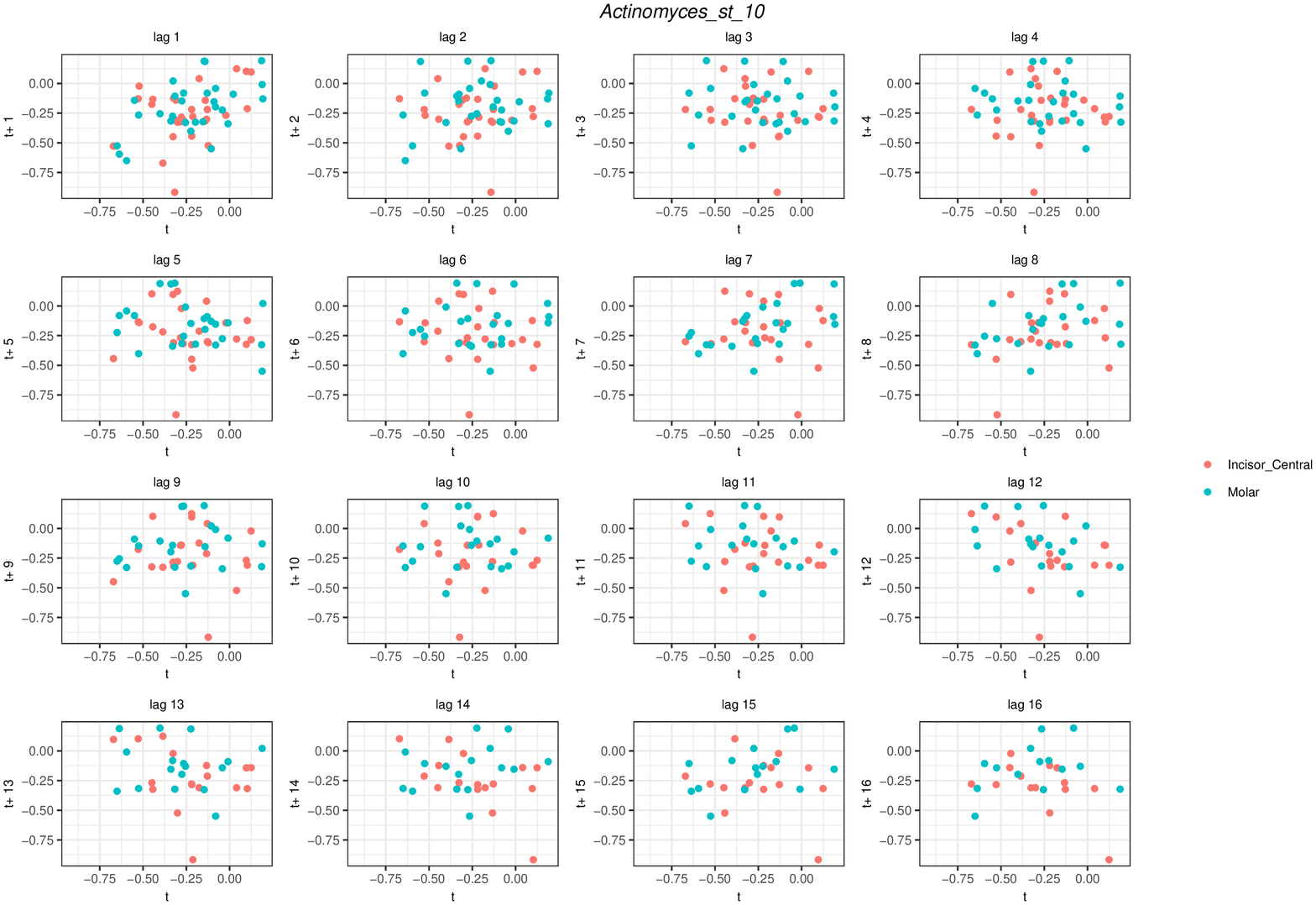} }  
	\end{tabular}
	\caption{ \label{fig16}\textbf{Top:} Each facet shows a partial autocorrelation (PAC) plot of different ASVs in oral data. Colors refer to the level of a group variable (molar/incisor). The x-axis label $h$ denotes the lag. The larger spikes (PAC $>$ .25) are observed at lags less than 10 for both molar and incisor tooth classes, except for large spikes at lag-12, lag-15, lag-16 in \textit{st\_17}, \textit{Streptococcus\_st\_2}, and \textit{Actinomyces\_st\_10}, respectively. We can check whether these are spurious PAC using lag-plots;  \textbf{Bottom:} For example, \textit{Actinomyces\_st\_10} lag-plot shows that the lag-16 observations that make cluster along the diagonal in negative direction are close to zero in incisor tooth class. Thus, those observations make a spurious effect. We can still choose $l_{I} = 11$.}
\end{figure}

\begin{figure}
	\centering
	\makebox{\includegraphics[width=\textwidth, scale = 1.5]{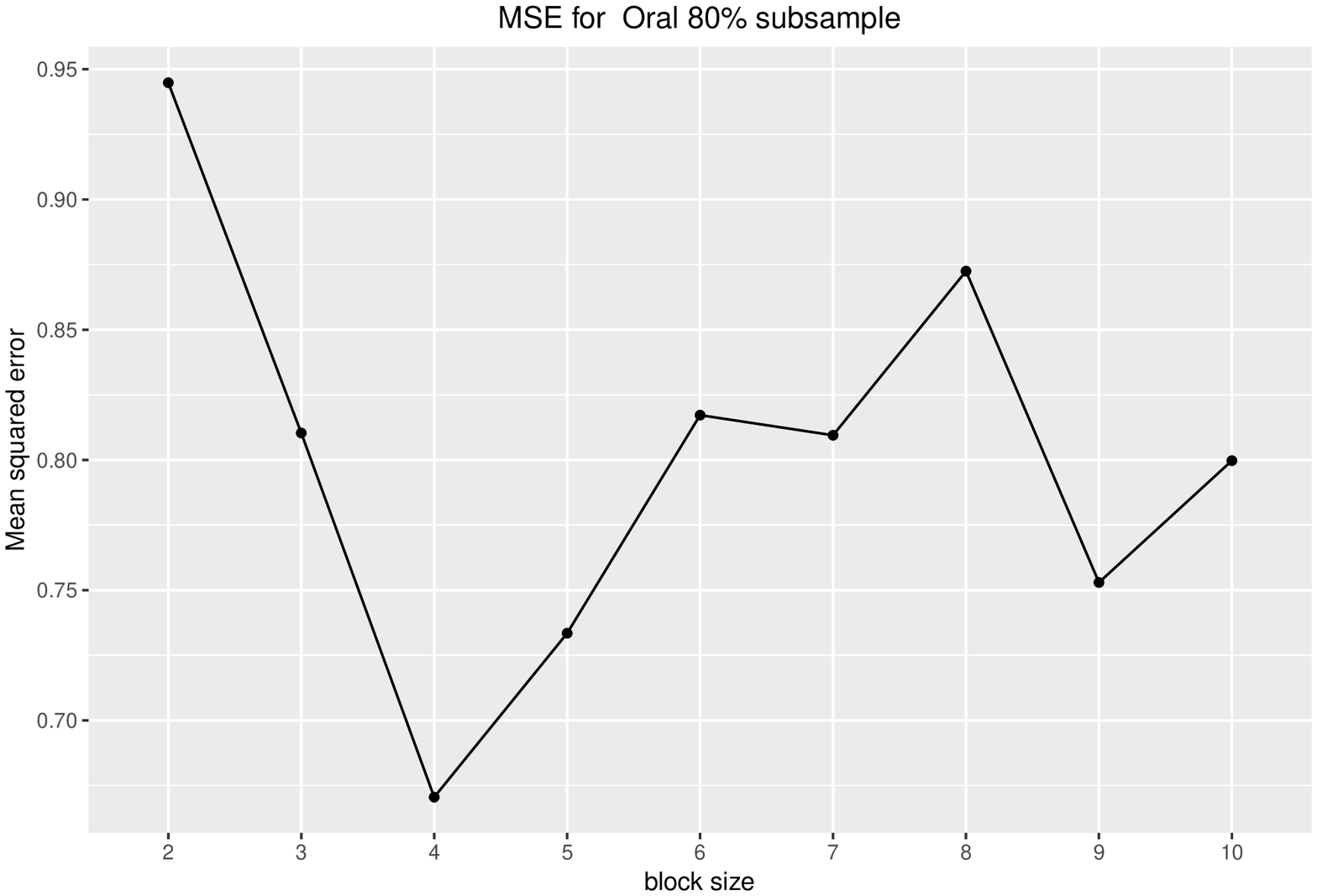}}
	\caption{\label{fig17} The minimum of mean squared error (MSE) occurs at block size four for $80\%$ subsample. The optimal block size for the oral data is four.}
\end{figure}

\begin{table}
	\caption{\label{tab12}Differentially abundant ASVs at FDR .05 in the oral data using the \textbf{MBB} method. ASV is identified with species/genus name and strain identification number, $\hat{\beta}_{i}$ is the fold change on the arcsinh scale of abundance in molar to incisors tooth classes, p.adj is the adjusted p-value, and CI is the 95\% confidence interval for $\beta_{i}$.
		}
	\centering
	\fbox{\begin{tabular}{llllll}
		 & Taxa &$\hat{\beta}_{i}$ & lcl & ucl & p.adj \\ 
		 \hline
	1 & Campylobacter\_concisus\_st\_55 & 2.66 & 2.47 & 3.02 & $<$.0001 \\ 
	2 & Prevotella\_denticola\_st\_74 & 2.62 & 2.12 & 3.06 & $<$.0001 \\ 
	3 & Prevotella\_histicola\_st\_36 & 2.56 & 2.39 & 2.89 & $<$.0001 \\ 
	4 & Veillonella\_st\_28 & 2.43 & 2.21 & 2.67 & $<$.0001 \\ 
	5 & Prevotella\_salivae\_st\_47 & 2.30 & 2.15 & 2.71 & $<$.0001 \\ 
	6 & Prevotella\_melaninogenica\_st\_37 & 2.27 & 1.97 & 2.69 & 0.0071 \\ 
	7 & Prevotella\_pallens\_st\_70 & 2.23 & 2.02 & 2.53 & $<$.0001 \\ 
	8 & Prevotella\_nigrescens\_st\_49 & 2.20 & 1.86 & 2.43 & $<$.0001 \\ 
	9 & Oribacterium\_sinus\_st\_71 & 2.18 & 1.99 & 2.52 & $<$.0001 \\ 
	10 & Prevotella\_nanceiensis\_st\_51 & 2.03 & 1.89 & 2.23 & $<$.0001 \\ 
	11 & Leptotrichia\_hongkongensis\_st\_25 & 2.00 & 1.78 & 2.33 & $<$.0001 \\ 
	12 & Prevotella\_st\_76 & 1.92 & 1.69 & 2.26 & $<$.0001 \\ 
	13 & Veillonella\_st\_75 & 1.63 & 1.47 & 2.01 & $<$.0001 \\ 
	14 & Prevotella\_oris\_st\_87 & 1.60 & 1.32 & 1.99 & $<$.0001 \\ 
	15 & Streptococcus\_st\_6 & 1.55 & 1.34 & 1.72 & $<$.0001 \\ 
	16 & Dialister\_invisus\_st\_78 & 1.46 & 1.03 & 1.82 & $<$.0001 \\ 
	17 & Solobacterium\_moorei\_st\_68 & 1.39 & 1.28 & 1.79 & 0.0071 \\ 
	18 & Veillonella\_st\_21 & 1.31 & 1.16 & 1.38 & $<$.0001 \\ 
	19 & Haemophilus\_st\_35 & 1.30 & 1.03 & 1.56 & $<$.0001 \\ 
	20 & Prevotella\_nanceiensis\_st\_42 & 1.28 & 0.94 & 1.67 & $<$.0001 \\ 
	21 & Mogibacterium\_st\_72 & 1.27 & 1.14 & 1.42 & $<$.0001 \\ 
	22 & Prevotella\_oris\_st\_88 & 1.23 & 0.91 & 1.52 & $<$.0001 \\ 
	23 & Porphyromonas\_st\_79 & 1.21 & 0.76 & 1.76 & 0.0457 \\ 
	24 & Veillonella\_st\_7 & 1.12 & 0.94 & 1.24 & $<$.0001 \\ 
	25 & Leptotrichia\_st\_65 & 1.11 & 0.82 & 1.59 & 0.0136 \\ 
	26 & Neisseria\_st\_15 & 1.06 & 0.91 & 1.22 & $<$.0001 \\ 
\vdots \\
	42 & Neisseria\_st\_4 & -1.06 & -1.18 & -0.97 & $<$.0001 \\ 
	43 & Actinomyces\_st\_10 & -1.18 & -1.37 & -1.02 & $<$.0001 \\ 
	44 & Actinomyces\_massiliensis\_st\_34 & -1.21 & -1.36 & -1.00 & $<$.0001 \\ 
	45 & Capnocytophaga\_sputigena\_st\_48 & -1.49 & -1.78 & -1.28 & $<$.0001 \\ 
	46 & Actinomyces\_st\_39 & -1.51 & -1.68 & -1.34 & $<$.0001 \\ 
	47 & Cardiobacterium\_hominis\_st\_24 & -1.59 & -1.89 & -1.33 & $<$.0001 \\ 
	48 & Streptococcus\_sanguinis\_st\_5 & -1.73 & -1.81 & -1.61 & $<$.0001 \\ 
	49 & Rothia\_dentocariosa\_st\_1 & -2.06 & -2.17 & -1.95 & $<$.0001 \\ 
	50 & Rothia\_st\_11 & -2.28 & -2.41 & -2.11 & $<$.0001 \\ 
	51 & Neisseria\_st\_26 & -2.45 & -2.72 & -2.27 & $<$.0001 \\ 
	52 & Corynebacterium\_durum\_st\_16 & -2.83 & -2.90 & -2.75 & $<$.0001 \\ 
		\end{tabular}}
	\end{table}
	
\begin{figure}
	\centering
	\makebox{\includegraphics[width=\textwidth, scale =2]{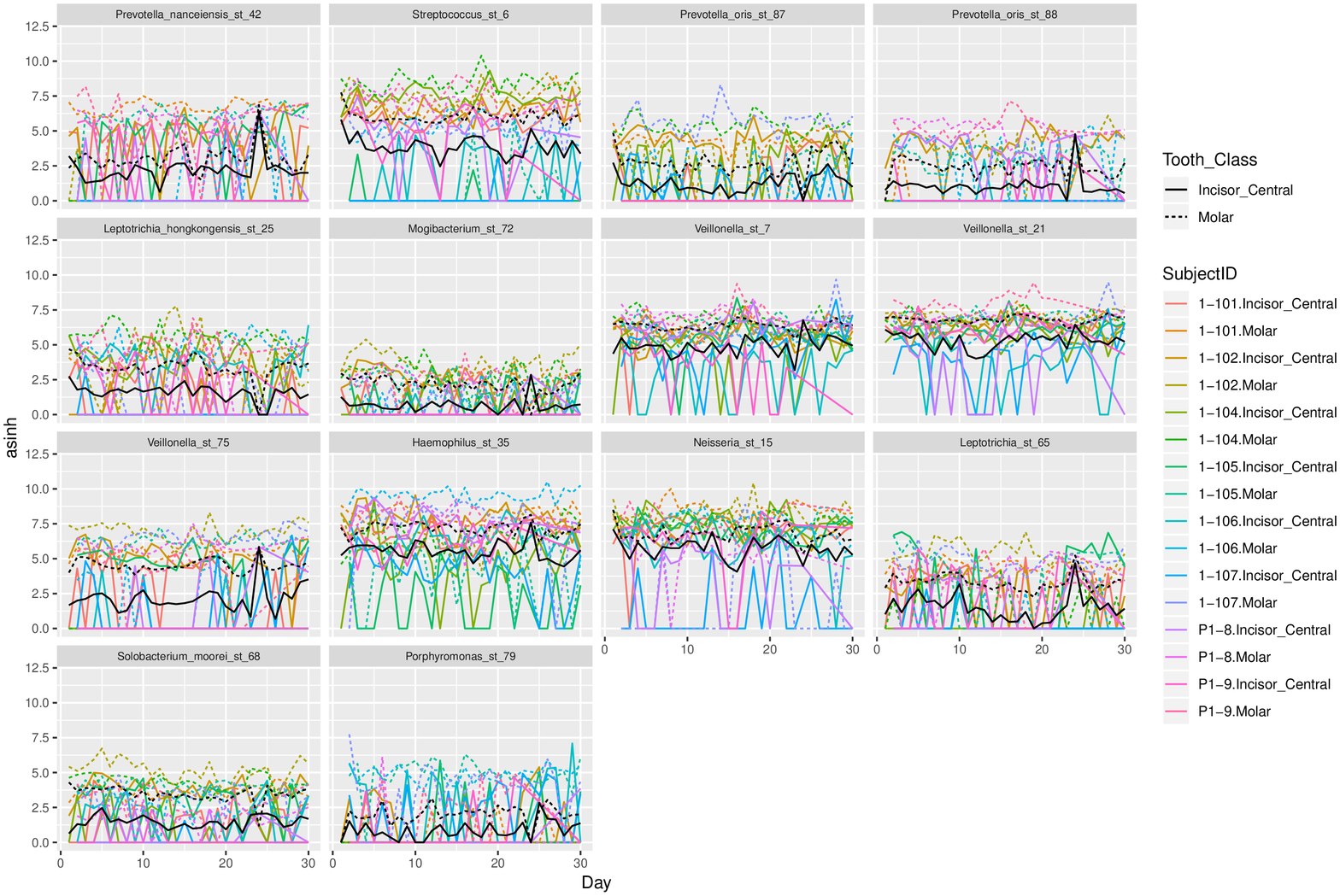}}
	\caption{\label{fig18} arcsinh transformed abundances of significant ASVs with the MBB method but not with MBS method. The black solid line and dotted lines are the mean transformed abundances over time in incisors and molars, respectively.}
\end{figure}
	
\begin{figure}
	\centering
	\makebox{\includegraphics[width=\textwidth, scale = 2]{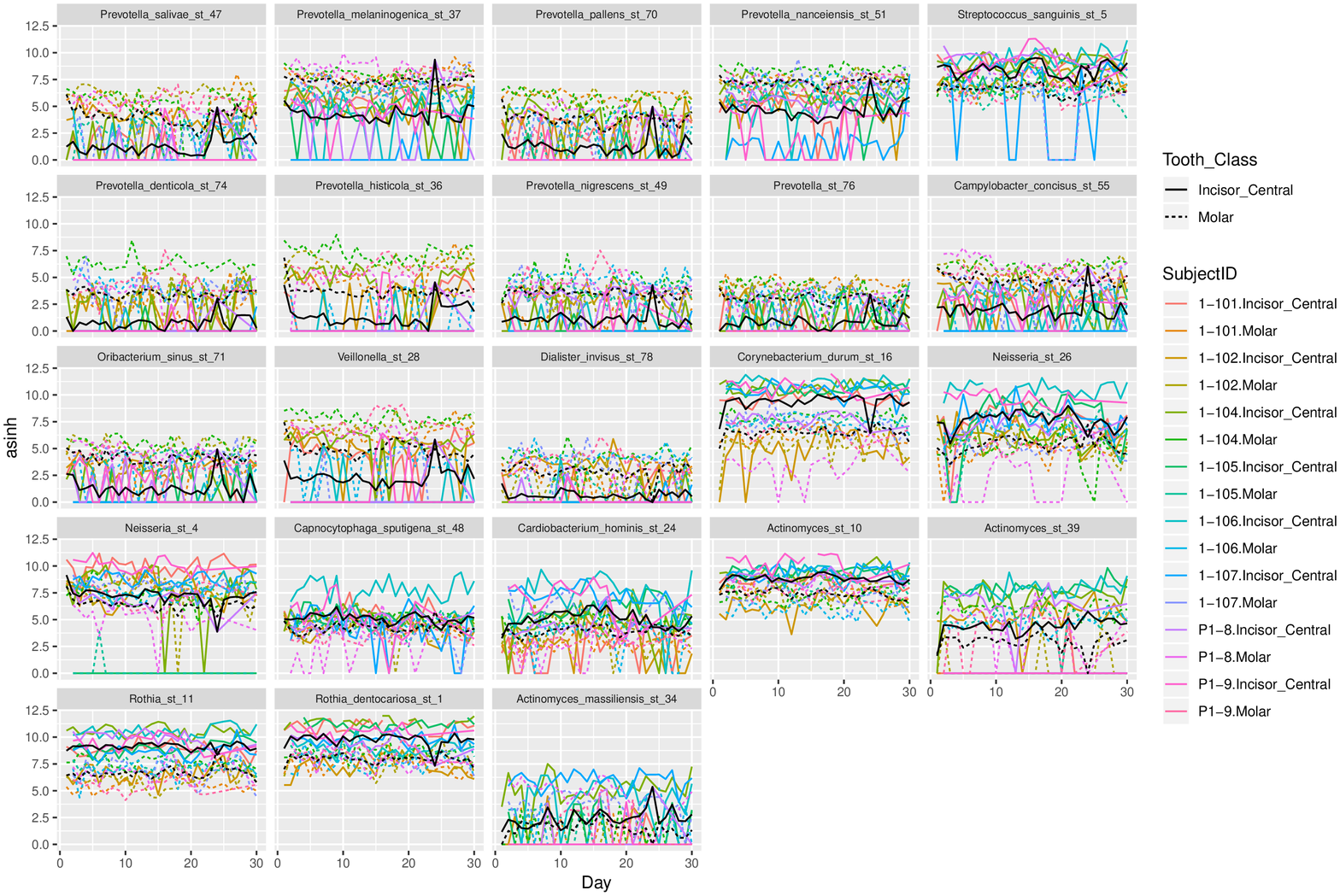}}
	\caption{\label{fig19} arcsinh transformed abundances of significant ASVs with the MBB and MBS method in oral data.}
\end{figure}

\end{document}